\documentclass[sigconf]{acmart}

\copyrightyear{2025}
\acmYear{2025}
\setcopyright{rightsretained}
\acmConference[FPGA '25]{Proceedings of the 2025 ACM/SIGDA International Symposium on Field Programmable Gate Arrays}{February 27--March 1, 2025}{Monterey, CA, USA}
\acmBooktitle{Proceedings of the 2025 ACM/SIGDA International Symposium on Field Programmable Gate Arrays (FPGA '25), February 27--March 1, 2025, Monterey, CA, USA}
\acmDOI{10.1145/3706628.3708867}
\acmISBN{979-8-4007-1396-5/25/02}

\usepackage{comment}
\usepackage{graphicx}
\usepackage{xcolor}
\usepackage{multirow}
\usepackage{siunitx}
\usepackage{todonotes}
\usepackage{hyperref}
\usepackage{amsmath}
\usepackage{amsfonts}
\usepackage{algorithmic}
\usepackage{subfig}
\usepackage{makecell}
\usepackage{svg}
\usepackage{tablefootnote}
\usepackage[flushleft]{threeparttable}
\usepackage[symbol]{footmisc}
\usepackage{fixltx2e}
\usepackage{caption} 
\usepackage{soul}

\usepackage{url}

\begin{document}

\title{Systolic Sparse Tensor Slices: FPGA Building Blocks for Sparse and Dense AI Acceleration}

\author{Endri Taka}
\affiliation{%
  \institution{The University of Texas at Austin}
  \city{Austin}
  \state{TX}
  \country{United States}
}
  \email{endri.taka@utexas.edu}

\author{Ning-Chi Huang}
\affiliation{%
  \institution{National Yang Ming Chiao Tung University}
  \city{Hsinchu}
  \country{Taiwan}
}

\author{Chi-Chih Chang}
\affiliation{%
  \institution{National Yang Ming Chiao Tung University}
  \city{Hsinchu}
  \country{Taiwan}
}

\author{Kai-Chiang Wu}
\affiliation{%
  \institution{National Yang Ming Chiao Tung University}
  \city{Hsinchu}
  \country{Taiwan}
}

\author{Aman Arora}
\affiliation{%
  \institution{Arizona State University}
  \city{Tempe}
  \state{AZ}
  \country{United States}
}

\author{Diana Marculescu}
\affiliation{%
  \institution{The University of Texas at Austin}
  \city{Austin}
  \state{TX}
  \country{United States}
}

\renewcommand{\shortauthors}{Endri Taka et al.}

\begin{abstract}

FPGA architectures have recently been enhanced to meet the substantial computational demands of modern deep neural networks (DNNs).
To this end, both FPGA vendors and academic researchers have proposed in-fabric blocks that perform efficient tensor computations. 
However, these blocks are primarily optimized for dense computation, while most DNNs exhibit sparsity. 
To address this limitation, we propose incorporating \textit{structured} sparsity support into FPGA architectures.
We architect 2D systolic in-fabric blocks, 
named systolic sparse tensor (SST) slices, that support multiple degrees of sparsity to efficiently accelerate a wide variety of DNNs.
SSTs support dense operation, 2:4 (50\%) and 1:4 (75\%) sparsity, as well as a new 1:3 (66.7\%) sparsity level to further increase flexibility.
When demonstrating on general matrix multiplication (GEMM) accelerators, which are the heart of most current DNN accelerators, our sparse SST-based designs attain up to 5$\times$ higher FPGA frequency and 10.9$\times$ lower area, compared to traditional FPGAs.
Moreover, evaluation of the proposed SSTs on state-of-the-art sparse ViT and CNN models exhibits up to 3.52$\times$ speedup with minimal area increase of up to 13.3\%, compared to dense in-fabric acceleration.

\end{abstract}




\maketitle

\section{Introduction}
\label{sec:Introduction}
Sparsity, which refers to zeros in weight or activation tensors, is an inherent attribute of contemporary deep neural networks (DNNs) \cite{Sparsity_Hoefler_2021}.
Sparsity arises from complex interactions among various  optimization techniques in modern DNN models \cite{HighLight_MIT_2023}.
For instance, over-parameterization is prevalent because it enables easier training and better generalization \cite{Sparsity_Hoefler_2021}.
Pruning is typically applied for this over-parameterization redundancy, resulting in zeros within the weight tensors (sparsity).
The exploitation of sparsity to reduce computational and memory requirements has been a common strategy in the design of efficient DNN accelerators  \cite{S2TA_HPCA_2022}.

Typically, zeros in DNNs are distributed randomly within the data tensors \cite{Sparsity_Hoefler_2021, S2TA_HPCA_2022}. 
This random sparsity is commonly referred to as \textit{unstructured} sparsity.
Leveraging unstructured sparsity has been the main focus of numerous sparse hardware accelerators over the past years, targeting both ASICs \cite{EIE_ISCA_2016, Cambricon_X_MICRO_2016, OuterSpace_HPCA_2018, SMT_SA_2019, ExTensor_2019, SparTen_2019, Eyeriss_v2_2019, SpArch_2020} as well as FPGAs \cite{unstructured_sparsity_SpWA_FPGA_DAC_2018,
unstructured_sparsity_CNN_FPGA_FCCM_2019,
unstructured_sparsity_FPGA_UIUC_2019,
unstructured_sparsity_FPGA_Abhishek_2021,
unstructured_sparsity_CNN_TCS_2021, unstructured_FPGA_TCAD_2022}.
However, unstructured sparse accelerators require complex hardware structures, which result in high area overheads \cite{HighLight_MIT_2023}. 
This area increase leads to substantial rise in energy consumption, \emph{e.g.,} up to 71\% \cite{S2TA_HPCA_2022}, compared to dense architectures.
Moreover, the random location of zeros in unstructured sparsity causes unpredictable and low hardware utilization, rendering inference speedup inefficient as well as challenging \cite{STA_arxiv_2020, STA_arch_letters_2020, EIE_ISCA_2016}.

To address these challenges in sparsity acceleration, 
\textit{structured} sparsity has been proposed recently \cite{Nvidia_accelerate_sparse_2021, S2TA_HPCA_2022, Vegeta_HPCA_2023}.
Structured sparsity imposes constraints on the sparsity patterns in data tensors, enabling low area overheads and highly energy-efficient hardware enhancements for sparsity exploitation.
To this end, Nvidia A100 GPUs introduced support for the 2:4 structured sparsity, 
where two out of every four \textit{consecutive} values must be non-zero.
Furthermore, structured sparse ASIC accelerators have also been proposed, demonstrating up to 3.1$\times$ less energy consumption per inference compared to state-of-the-art unstructured sparse accelerators \cite{S2TA_HPCA_2022}.

Alongside GPUs and ASICs, FPGAs have emerged as a promising candidate for accelerating the rapidly evolving DNNs, due to their high flexibility of reconfiguration.
Contemporary FPGA architectures have been enhanced to more  
effectively support the high computational demands of DNNs.
In particular, Intel employs in-fabric blocks comprising multiple dot-product engines \cite{Stratix_10_NX_FPGA_2021, Sergey_Intel_TB_Agilex_5_FCCM_2024}, while AMD introduced out-of-fabric programmable vector processors \cite{AMD_AIE_ML_architecture_manual}.
Academic researchers have also proposed in-fabric blocks that employ a 2D systolic dataflow \cite{TS_Aman_FPGA_2021, Aman_TS_TRETS_2022}, exhibiting substantial benefits over traditional FPGAs for various DNN workloads.

However, these in-fabric blocks are primarily designed for dense DNN acceleration, limiting their efficiency and applicability to most real-world DNNs, which inherently exhibit sparsity.
To address this challenge, we propose incorporating flexible in-fabric slices into the FPGA architecture to efficiently support both \textit{sparse} and \textit{dense} DNN workloads.
We employ 2D systolic slices similar to \cite{TS_Aman_FPGA_2021, Aman_TS_TRETS_2022}, which are augmented with structured sparsity features and further optimizations.
Our novel systolic sparse tensor (SST) slices are architected with the following principal properties: 

\vspace{-0.10cm}
\begin{itemize}
  \item \textbf{Efficiency}. (\textbf{i}) Sparsity features should exhibit low area overheads. (\textbf{ii}) Translation of sparsity to actual DNN acceleration. (\textbf{iii}) Maximization of data reuse in sparse and dense DNNs. (\textbf{iv}) Efficient sparse format for storing the non-zero values. 
  \item \textbf{Sparsity Flexibility}. These in-fabric blocks must be flexible, \emph{i.e.,} support multiple structured sparsity levels (percentage of zeros) to accelerate a wide range of DNN workloads.
\end{itemize}
\vspace{-0.10cm}

The SST slices are systolic-based and utilize a highly efficient index-based sparse format, effectively meeting the aforementioned key properties for \textit{efficiency}.
Regarding the \textit{sparsity flexibility}, SSTs support multiple sparsity levels, \emph{i.e.,} dense, 2:4 (50\% sparsity), 1:3 (66.7\%) and 1:4 (75\%).
These levels align with the most \textit{common} sparsity levels in DNNs, since sparsity higher than 75\% typically leads to significant accuracy degradation \cite{Sparse_tensor_GPUs_2019, STA_arxiv_2020, N_M_sparse_transformers_FPGA_VLSI_2022, Learning_N_M_ICLR_2021, Learning_best_N_M_NeurIPS_2022}.
Prior works have exploited the 2:4 and 1:4 sparsity patterns to design sparse accelerators using traditional FPGAs \cite{LAMPS_FCCM_2024, N_M_sparse_transformers_FPGA_VLSI_2022}, while others have incorporated these patterns into CPU datapaths \cite{Vegeta_HPCA_2023, RISC_V_CPU_structured_sparsity_DATE_2024}.
In contrast, we propose a novel in-fabric FPGA block that additionally supports a new 1:3 sparsity pattern.
The 
1:3 sparsity bridges the gap between 50\% and 75\% sparsity, increasing flexibility and providing both acceleration and storage benefits.
Furthermore, 1:3 sparsity can be supported in SSTs 
\textit{without incurring additional area overheads},
by effectively reusing the area allocated for 2:4 and 1:4.
This sparsity flexibility enables the development of tailored solutions for DNN models, since each model exhibits distinct levels of sparsity.

The proposed SST slices support 8-bit integer (int8) and brain floating-point (bfloat16) precisions, since these are the most commonly used data types in DNN accelerators \cite{TPUv42021, TPUV2_v3_2021}.
Additionally, we introduce \textit{dedicated} interconnects between the SST slices to facilitate their efficient integration within the FPGA architecture.
These dedicated interconnects demonstrate substantial routing savings in FPGAs compared to prior work on 2D systolic blocks \cite{TS_Aman_FPGA_2021, Aman_TS_TRETS_2022}, where such interconnects are not employed.

The SST slices deliver highly efficient acceleration of sparse and dense general matrix multiplication (GEMM) in the FPGA fabric.
Our main focus is GEMM, as most current state-of-the-art DNNs across various applications are Transformer-based \cite{Transformers_SOTA_2023, Transformer_SOTA_Elsevier_2024}, with GEMM serving as the core computation. 
To the best of our knowledge, this is the first work to support structured sparsity in the FPGA architecture.
Our key contributions are summarized below:

\vspace{-0.10cm}
\begin{itemize}
  \item  A \textit{generalizable methodology} for incorporating structured sparsity into in-fabric FPGA blocks. Our proposed SST slices employ a 2D systolic dataflow and support multiple levels of sparsity. Besides dense, 2:4 and 1:4, we also introduce a \textit{new 1:3 sparsity pattern} to provide greater flexibility, enabling efficient acceleration for the majority of DNN models.
  \item  We introduce dedicated interconnects to effectively integrate the SST slices in the FPGA architecture. These interconnects show \textit{significant routing savings, up to \textbf{31.2\%}} on GEMM accelerators, compared to utilizing only global routing resources, as proposed in prior work on 2D systolic blocks.
  \item Demonstration on sparse GEMM accelerators leveraging our SST slices show considerably \textit{higher attainable FPGA frequency, up to \textbf{4.4$\times$} for int8 and \textbf{5$\times$} for bfloat16}, as well as \textit{remarkable area reduction, up to \textbf{7$\times$} for int8 and \textbf{10.9$\times$} for bfloat16}, when comparing with traditional FPGAs.
  \item Our evaluation on state-of-the-art sparse ViT and CNN models showcases up to \textit{\textbf{3.52$\times$} speedup} when exploiting our SST slices compared to dense in-fabric blocks, with \textit{low area overheads of \textbf{10.2\%} and \textbf{13.3\%} for int8 and bfloat16}, respectively.
\end{itemize}
\vspace{-0.20cm}

\section{Related Work}
\label{sec:Related_work}

Structured sparsity is currently supported by various commercial architectures.
In particular, the 2:4 sparsity is supported by state-of-the-art GPUs, \emph{i.e.,} Nvidia A100 \cite{NVIDIA_A100}, Nvidia H100 \cite{NVIDIA_H100} and AMD MI300 \cite{AMD_CDNA_3}, as well as the new AMD Versal AIE-ML 
processors \cite{AMD_AIE_ML_architecture_manual, AMD_AIE_ML_kernel_guide}.
In academic research, hardware support for structured sparsity has also been investigated in ASIC DNN accelerators \cite{S2TA_HPCA_2022, STA_arch_letters_2020, HighLight_MIT_2023}.
Moreover, prior academic research \cite{Sparse_tensor_GPUs_2019} introduces structured sparsity features in the tensor cores of modern GPUs.
Finally, structured sparsity support has also been proposed inside the matrix engines \cite{Vegeta_HPCA_2023} and vector units \cite{RISC_V_CPU_structured_sparsity_DATE_2024} of CPUs.
Considering all prior work, we are the first to incorporate structured sparsity support into the FPGA architecture.
Moreover, our proposed in-fabric blocks support multiple levels of structured sparsity (beyond merely 2:4), to enable efficient acceleration for most contemporary DNN models.

In-fabric hard blocks have been commercially incorporated into several AI-optimized FPGAs.
More specifically, the Intel Stratix 10 NX \cite{Stratix_10_NX_FPGA_2021} replaces the traditional DSP slices with tensor blocks.
These tensor blocks comprise multiple dot-product engines to more efficiently support AI applications.
Very recently, Intel announced the Agilex-5 FPGAs \cite{Sergey_Intel_TB_Agilex_5_FCCM_2024, Intel_Agilex_5_tensor_blocks}, which introduce new AI-enhanced DSP blocks.
These new DSPs include multiple dot-product operations, while also maintaining various features of the traditional Agilex DSP blocks.
Finally, the Achronix Speedster7t FPGAs incorporate machine learning processor (MLP) blocks \cite{Achronix_Speedster, Achronix_Speedster_2024}.
These MLP hard blocks feature multiplier arrays, adder trees, accumulators and tightly coupled memories to the computational blocks.

In-fabric enhancements have also been proposed in academia.
In \cite{TS_Aman_FPGA_2021, Aman_TS_TRETS_2022}, the authors enrich the FPGA architecture with 2D systolic tensor slices, showing substantial efficiency benefits in DNN acceleration over traditional FPGAs.
Moreover, in \cite{PIR_DSP_FCCM_2019}, a special pattern of dedicated wires between DSP blocks is proposed, which enables more efficient mapping of systolic arrays on FPGAs.
However, all aforementioned commercial and academic in-fabric blocks target primarily dense computation.
In this work, we architect in-fabric slices that enable both sparse and dense computation.
We propose incorporating the SST slices, which employ a 2D systolic dataflow similar to \cite{TS_Aman_FPGA_2021, Aman_TS_TRETS_2022}, but are augmented with sparsity features.
Moreover, we introduce vertical dedicated wires between the SSTs to more efficiently map GEMM accelerators into FPGA architectures.

Finally, prior research \cite{LSTM_fine_grained_sparsity_FPGA_2019, Fine_grained_structured_sparsity_FPL_2021, N_M_sparse_transformers_FPGA_VLSI_2022, Fine_grained_Neural_ODE_FPGA_2023, LAMPS_FCCM_2024} implement DNN accelerators that support multi-level structured sparsity on traditional (non AI-optimized) FPGAs. 
In contrast, we also introduce the novel 1:3 sparsity pattern, which is absent from any prior work and offers increased sparsity flexibility.
In this work, we demonstrate the importance of this flexibility for various state-of-the-art DNNs.
Furthermore, our SST-based GEMM designs show remarkable performance and area efficiency gains compared to traditional FPGAs.

\section{Architecture \& Design Overview}
\label{sec:Architecture_Overview}

\subsection{Fine-Grained Structured Sparsity}
\label{subsec:Fine_grained_structured_sparsity}
\begin{figure}[t]
\vspace{-0.70cm}
\centering
\subfloat[]
{\includegraphics[width=0.31\linewidth]{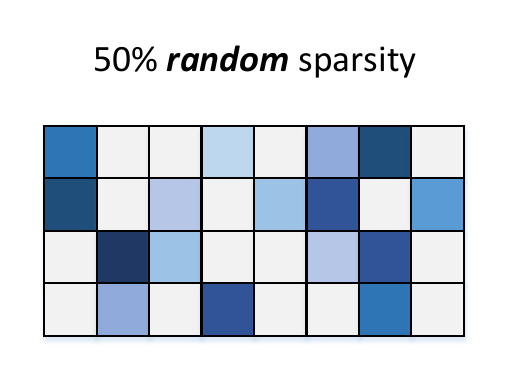}
\label{fig:random_sparsity}}
\subfloat[]{\includegraphics[width=0.685\linewidth]{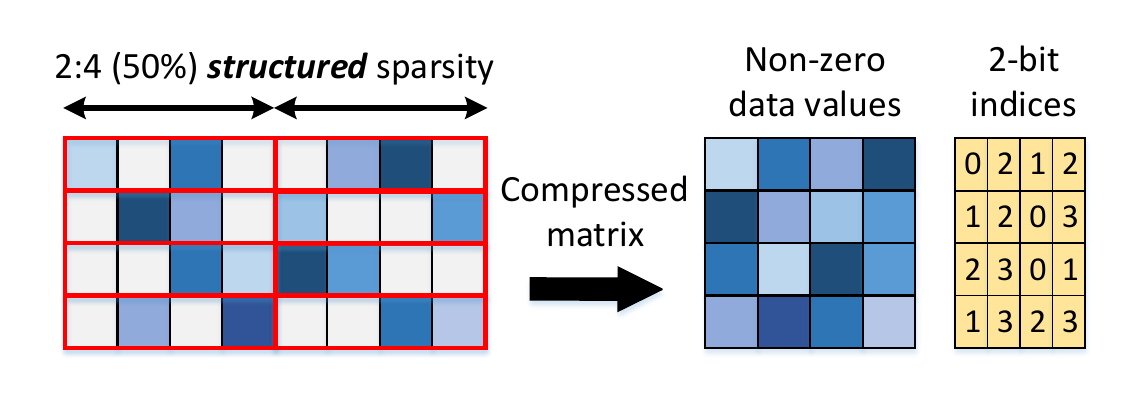}
\label{fig:structured_sparsity}} 

\vspace{-0.35cm}

\caption{50\% unstructured sparse matrix (a) and 2:4 (50\%) structured sparse matrix along with compressed format (b).} 
\label{fig:sparsity_struct_unstruct}
\vspace{-0.50cm}
\end{figure}

In this work, we leverage the regular patterns of fine-grained structured sparsity to architect SST slices with low area overhead. 
Fig. \ref{fig:random_sparsity} depicts a 50\% unstructured sparse matrix, where the non-zero data are distributed \textit{randomly}, \emph{i.e.,} there is no specific pattern of their locations.
In contrast, in Fig. \ref{fig:structured_sparsity}, the 2:4 structured sparsity pattern is illustrated, which has the same sparsity level (50\%), but in every group of four \textit{consecutive} elements
there are two non-zero values. 
Notice that the location of the two non-zero values can vary significantly within the four-element group, offering \textit{fine-grained} sparsity flexibility.  
These types of constraints enable low area hardware enhancements, allowing for efficient  exploitation of sparsity.

A 2:4 sparse matrix can be efficiently stored in  compressed format by saving only the non-zero values.
The location of each non-zero data is encoded using 2-bit indices, as shown in Fig. \ref{fig:structured_sparsity}.
Similar to 2:4, for 1:4 (75\%) sparsity one every four consecutive elements is non-zero, while for 1:3 (66.7\%) sparsity there is one non-zero every three consecutive elements.  
For all aforementioned patterns, 2-bit indices are required to encode the location of each non-zero element.
This is a very efficient compressed format, as we show in Sec. \ref{subsec:AIE_ML_comparison}, where comparison among other formats is performed.

\subsection{Systolic Sparse Tensor Slices Architecture}
\label{subsec:Sparse_Tensor_slices_architecture}

\begin{figure}[tbp]
\vspace{-0.50cm}
\centering
\includegraphics[width=0.89\linewidth]{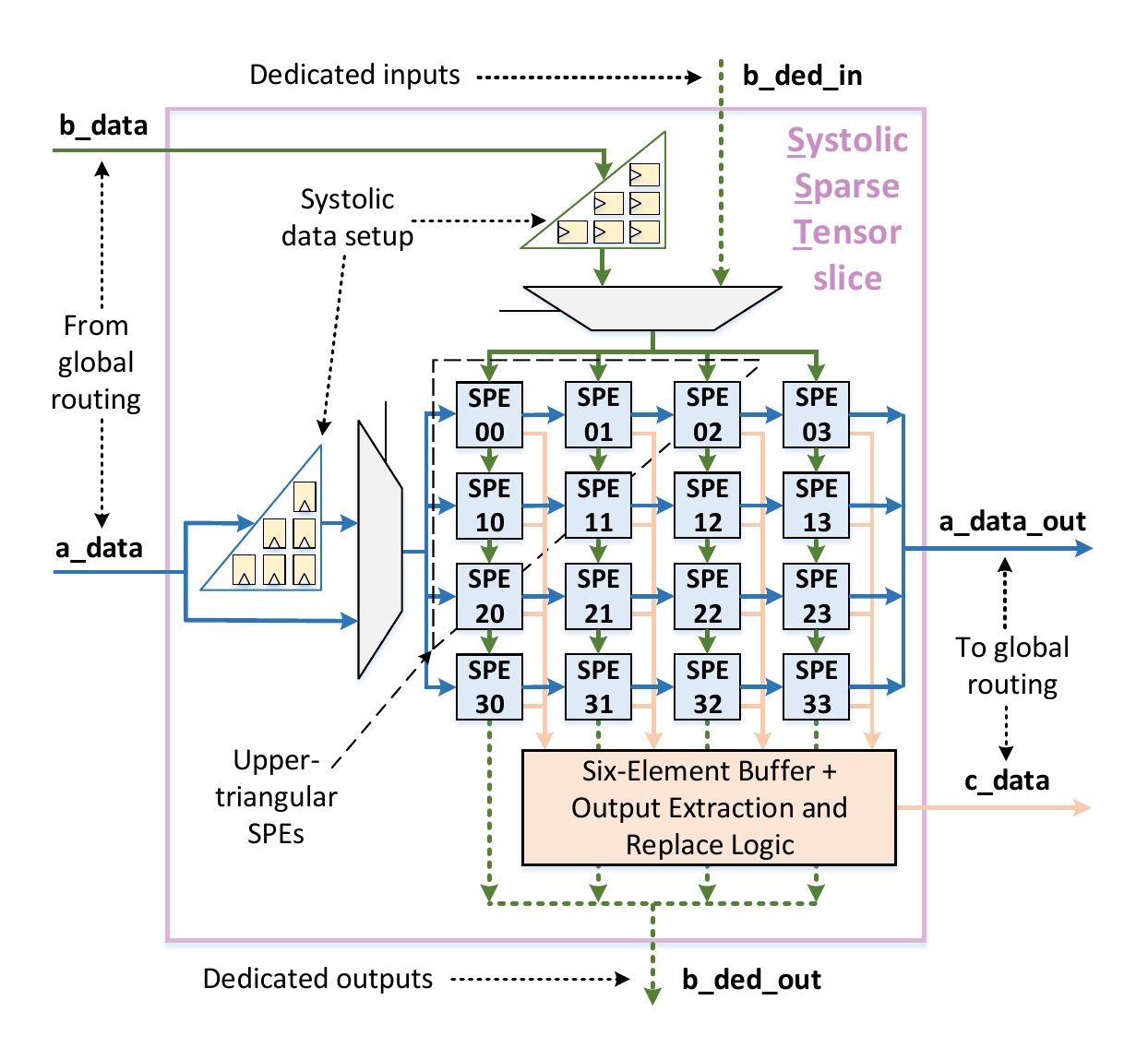}

\vspace{-0.50cm}

\caption{Systolic Sparse Tensor slice architecture.}
\label{fig:ST_architecture}

\vspace{-0.60cm}

\end{figure}

In this section, we present an
overview of the proposed SST slices. 
The core compute unit of the SSTs is a 4$\times$4 systolic array (SA) \cite{Kung_SA_1982}, as depicted in Fig. \ref{fig:ST_architecture}.
A 2D SA consists of homogeneous processing elements (PEs), where each PE performs a multiply--accumulate (MAC) operation and forwards the input operands to the neighboring PEs. 
This architecture allows maximization of data reuse
in GEMM, 
while also delivering high performance due to its regular and highly scalable design.
Hence, SAs have become a prime architecture in many DNN accelerators \cite{TPUV2_v3_2021, TPUv42021, S2TA_HPCA_2022, Vegeta_HPCA_2023, SA_CNN_FPGA_2017, SA_attention_FPGA_TECS_2023, Scale_sim_2020}.
In this work, we show that incorporating SST slices in FPGAs leads to high performance and scalable dense/sparse GEMM accelerators.

\subsubsection{SST Operation} 
We enhance the systolic PEs with sparse features by introducing sparse processing elements (SPEs), while maintaining the properties of the SAs discussed above. 
Our SSTs utilize an output stationary SA, consisting of 16 SPEs.
The accumulations remain stationary in the SPEs, while input operands are propagated to their neighbors every clock cycle.
The $a\_data$ of an input matrix $A$ are propagated and reused across SPEs horizontally, while the $b\_data$ of an input matrix $B$ are propagated and reused vertically (Fig. \ref{fig:ST_architecture}).
The matrix $A$ can be either sparse or dense (typically to map \textit{weights}), while $B$ is dense (typically to map input \textit{activations}).
The architecture of the SPEs is delineated in Sec. \ref{subsec:Sparse_Processing_Element}.

Besides the 4$\times$4 SPE grid, we also implement pipeline registers to delay the input operands for systolic data setup \cite{TPUv1_2017} (arranged in triangular manner in Fig. \ref{fig:ST_architecture}).
Systolic setup is needed at the interface for loading input matrices (typically from on-chip memory), when chaining multiple SSTs to construct larger SA grids (Sec. \ref{subsec:Matrix_multiplication_mapping}).
Multiplexers are used to select either the systolic setup
or directly the input data, and are configured \textit{statically} (during bitstream loading).

\begin{figure*}[tbp]
\vspace{-0.40cm}
\centering
\includegraphics[width=1.00\textwidth]{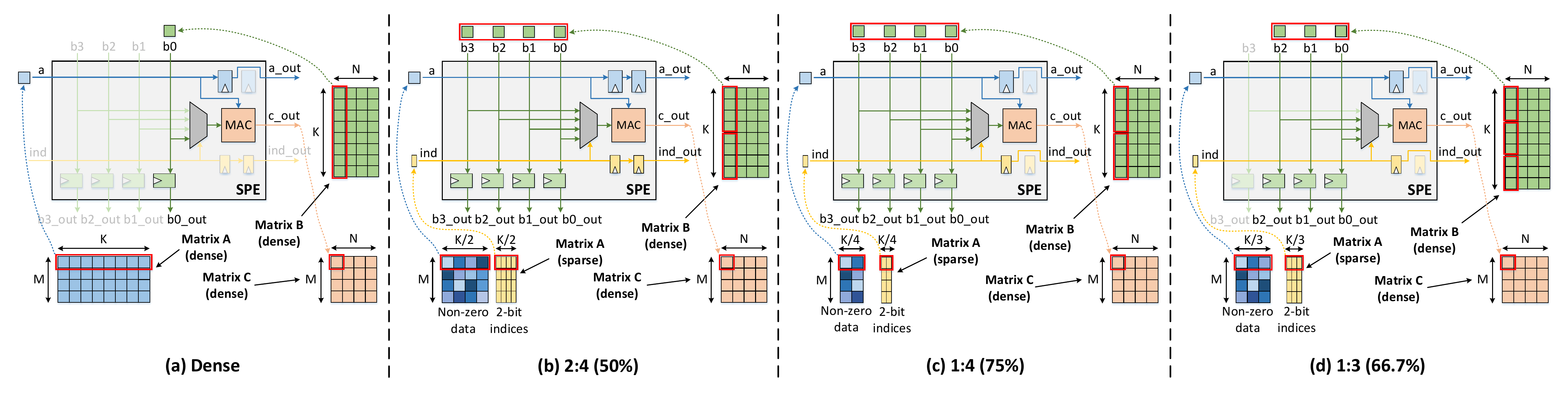}

\vspace{-0.45cm}

\caption{Sparsity modes in Systolic Sparse Elements of the SST slices (multiplexing logic omitted for clarity).}
\label{fig:SPE_sparse_modes}
\vspace{-0.35cm}
\end{figure*}

The SSTs support both int8 and bfloat16 precisions, while accumulations are realized in 32-bit integer (int32) and IEEE 32-bit floating-point (fp32), respectively, similar to Nvidia GPUs \cite{Nvidia_accelerate_sparse_2021} and Google TPUs \cite{TPUV2_v3_2021}.
When the SA operation is completed, the output matrix $C$ is extracted via the $c\_data$ output ports.
We note that SPEs finish their operation in a \textit{diagonal} fashion cycle after cycle.
In the first cycle, $SPE00$ finishes, in the second cycle both $SPE01$ and $SPE10$ finish, etc. (Fig. \ref{fig:ST_architecture}). 
However, to maintain \textit{regularity}, thus simplifying the downstream logic (typically in CLBs), we extract the output values in a column-wise manner (four values per cycle).
To achieve that, we introduce a six-element buffer (consisting of registers), to store the data before getting extracted.
This is particularly important in sustaining 100\% SPE utilization (16 MACs per cycle) at the steady-state (matrices processed one after the other).

Since SPEs complete their operation diagonally, the six upper-triangular SPE outputs, shown in Fig. \ref{fig:ST_architecture}, need to be stored in the buffer.
Suppose a cycle $T$ where the outputs of the biggest diagonal, \emph{i.e.,} $SPEs$ $\{03, 12, 21, 30\}$ are generated.
In cycle $T$, the values of the first column, \emph{i.e.,} $SPEs$ $\{00, 10, 20, 30\}$ can be extracted, where $SPEs$ $\{00, 10, 20\}$ are loaded from the buffer, while only $SPE30$ is directly extracted.
In cycle $T + 1$, the values of $SPEs$ $\{03, 12, 21\}$ replace the position of $SPEs$ $\{00, 10, 20\}$ in the buffer (since they have been extracted).
Therefore, the locations of the six-element buffer are being effectively reused, and in cycle $T+1$, the second column can be extracted, \emph{i.e.,} $SPEs$ $\{01, 11, 21, 31\}$.
In a similar fashion, the rest two columns are extracted in cycles $T+2$ and $T+3$, respectively, while the buffer locations are efficiently reused due to replacement.

\subsubsection{Global Routing Interface \& Dedicated Wires}

As illustrated in Fig. \ref{fig:ST_architecture}, the $a\_data$ and $b\_data$ input ports as well as $c\_data$ output ports are connected to the global FPGA routing resources.
When chaining multiple SSTs, $a\_data$ and $b\_data$ are forwarded to their next in the chain SSTs, horizontally and vertically, respectively.
Horizontally, the data are propagated via the $a\_data\_out$ using the FPGA routing resources.
However, vertically, we utilize \textit{dedicated} wires to propagate the data (via the $b\_ded\_out$ ports) and connect them to the next SST in the \textit{same} FPGA column (via the $b\_ded\_in$ ports).
These vertical dedicated wires provide efficient connections without the usage of the global routing resources, matching the columnar nature of the modern FPGA fabric \cite{FPGA_architecture_2021, FPGA_for_DL_2024}.
This approach significantly reduces routing resources, as opposed to  \cite{TS_Aman_FPGA_2021, Aman_TS_TRETS_2022}, where all inputs/outputs (I/Os) of the in-fabric tensor slices are connected to global routing (see comparison in Sec. \ref{subsec:Dedicated_wires_benefits}).

Besides the I/Os shown in Fig. \ref{fig:ST_architecture}, the SSTs include also several dynamic control signals.
In particular, an input $enable$ signal is used to control the operation of the SSTs.
This signal should be deasserted when the operation of the SST needs to be stalled.
Moreover, an $accumulate$ signal is utilized to control the accumulation in the SPEs.
This signal remains asserted when accumulation is desirable in the SPEs, \emph{e.g.,} during tiling to process larger matrices.
However, for every new matrix operation the $accumulate$ signal should be deasserted for one cycle.
Another input signal is the  $d\_type$, which  dynamically selects the precision, \emph{i.e.,} int8 or bfloat16.
Finally, a 2-bit $sparsity\_level$ signal is employed to dynamically select the sparsity level of the matrix $A$, \emph{i.e.,} dense, 2:4, 1:3, 1:4.
This signal is utilized internally in the SSTs to control the operation of the supported sparsity modes, as discussed in the next section.

Regarding the outputs of the SST, an $accumulate\_out$ signal is used for systolic distribution of the $accumulate$ signal when chaining multiple SSTs.
This systolic distribution makes the control logic significantly simpler, since the $accumulate$ is set only for the first SST in the 2D array layout, while being distributed to the remaining SSTs, as explained in Sec. \ref{subsec:Matrix_multiplication_mapping}.
Another output signal is the $valid\_out$, which is asserted when the output $c\_data$ are extracted column-wise (Fig. \ref{fig:ST_architecture}). 
This signal simplifies the downstream logic, since only $valid\_out$ needs to be checked for valid output data.

\subsection{Sparse Processing Element}
\label{subsec:Sparse_Processing_Element}

The SPE architecture in all supported sparsity modes is illustrated in Fig. \ref{fig:SPE_sparse_modes}.
When the SPE is configured in dense mode (Fig. \ref{fig:SPE_sparse_modes}a), it operates as a regular dense PE.
In particular, one MAC operation is performed every cycle, while the elements of matrices $A$ and $B$ are forwarded to their neighboring SPEs (via pipeline registers) horizontally and vertically, respectively.
The output value of matrix $C$ in each SPE is calculated after $K$ cycles, when considering the $M$$\times$$K$$\times$$N$ matrix dimensions depicted in Fig. \ref{fig:SPE_sparse_modes}a.

In 2:4 mode (50\% sparsity), matrix $A$ is stored in a compressed format of $M$$\times$$K/2$ size for both values and the 2-bit indices, as shown in Fig. \ref{fig:SPE_sparse_modes}b.
In order to achieve speedup over the dense case, we increase the number of ports of each SPE in the vertical dimension, to load four elements of the matrix $B$ in parallel.
Furthermore, the 2-bit index is also loaded in the SPE to select the corresponding $B$ values (via a 4:1 multiplexer), which need to get multiplied with each $A$ value.
However, in 2:4 sparsity, for every four values of the matrix $B$, two MAC operations need to be performed.
Hence, the four $B$ values remain in the SPE registers for two clock cycles (the four green registers in Fig. \ref{fig:SPE_sparse_modes}b are loaded every two cycles).
To correctly synchronize the systolic operation, two pipeline stages are required for both the $A$ values and their indices (blue and yellow registers in Fig. \ref{fig:SPE_sparse_modes}b), which are loaded every cycle. 
In this manner, one MAC operation is performed every cycle, ensuring 100\% utilization.
Moreover, the output value of matrix $C$ in each SPE is calculated in $K/2$ cycles, achieving 2$\times$ acceleration over dense operation.

\begin{table}[t]
\centering
\caption{Summary of supported sparsity levels in SST slices.}

\setlength\tabcolsep{9pt}
\renewcommand{\arraystretch}{1.0}
\resizebox{0.80\linewidth}{!}{
\begin{tabular}{c|cc|c|c}
\Xhline{2.5\arrayrulewidth}

\textbf{Sparsity}  &  
\multicolumn{2}{c|}{\textbf{Compres. ratio}} & \textbf{Speedup} & \textbf{SPE} \\

\cline{2-3}

\textbf{level}  & \textbf{int8} & \textbf{bfloat16} 
 & \textbf{over dense} & \textbf{util.} \\

\hline
\hline

\textbf{Dense (0\%)} & 1$\times$ & 1$\times$ & 1$\times$ & 100\% \\ 

\textbf{2:4 (50\%)} & 1.6$\times$ & 1.78$\times$ & 2$\times$ & 100\% \\ 

\textbf{1:3 (66.7\%)} & 2.4$\times$ & 2.67$\times$ & 3$\times$ & 100\% \\ 

\textbf{1:4 (75\%)} & 3.2$\times$ & 3.56$\times$ & 4$\times$ & 100\% \\

\Xhline{2.5\arrayrulewidth}

\end{tabular}
}

\label{tb:sparsity_summary_benefits}

\vspace{-0.50cm}

\end{table}

\begin{figure*}[ht]

\vspace{-0.55cm}

\centering
\includegraphics[width=0.81\textwidth]{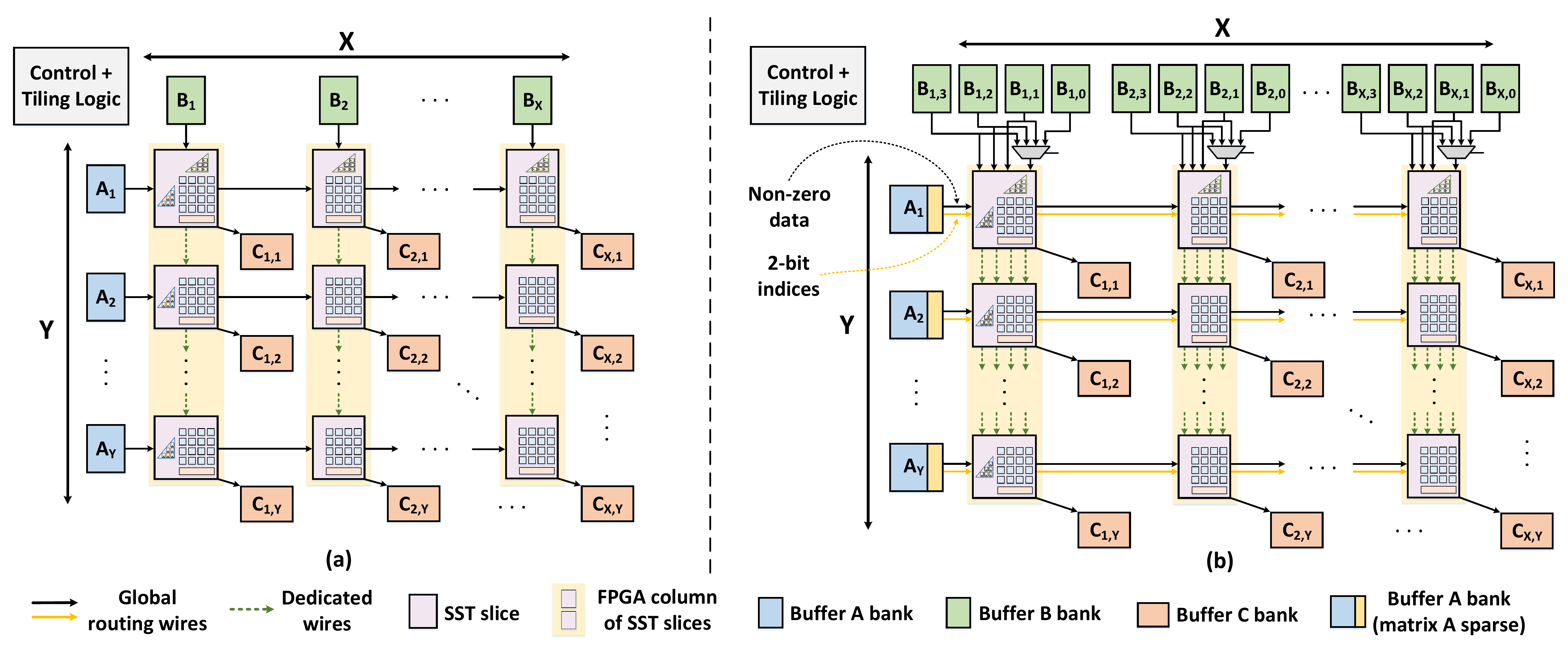}

\vspace{-0.35cm}

\caption{2D systolic GEMM design: dense implementation (a) and dynamic configuration of all supported sparsity modes (b).}
\label{fig:GEMM_2D_array_SSTs}

\vspace{-0.45cm}

\end{figure*}

Similar to 2:4, for 1:4 (75\%) sparsity, four $B$ values are loaded in the SPE.
However, in this case the $B$ values are loaded every cycle, while only one pipeline stage is required for the $A$ value and its index, as shown in Fig. \ref{fig:SPE_sparse_modes}c.
The output value needs $K/4$ cycles to be calculated, offering 4$\times$ speedup over the dense case.
To fill the sparsity gap between 50\% and 75\% as explained in Sec. \ref{sec:Introduction}, we leverage the same hardware enhancements for 2:4 and 1:4 sparsity to also support the 1:3 (66.7\%) pattern.
In particular, as illustrated in Fig. \ref{fig:SPE_sparse_modes}d, the 1:3 operation is similar to 1:4, with the main difference being that three $B$ values are loaded every cycle instead of four.
The indices ensure that the fourth $B$ value is never selected, thus no additional hardware is required.
In this case, $K/3$ cycles are needed for every SPE output value, providing 3$\times$ acceleration.
Finally, we note that similar to 1:3, 2:3 sparsity can also be supported by directly utilizing the 2:4 mode, leading to 33.3\% sparsity.
However, a matrix is typically considered sparse when it has sparsity of 50\% or higher \cite{AMD_AIE_ML_kernel_guide}.
Hence, in this work, we do not consider the 2:3 pattern.

In Table \ref{tb:sparsity_summary_benefits}, we present a summary of the supported sparsity levels.
First, we show the \textit{compression ratio}, \emph{i.e.,} the memory reduction over dense storage due to compressed format, for both int8 and bfloat16 precisions.
For all sparsity levels, bfloat16 offers a higher compression ratio over int8, \emph{e.g.,} 1.78$\times$ \emph{vs.} 1.6$\times$ for 2:4 sparsity. 
This is because 2-bit indices are required for both 8-bit and 16-bit data types, resulting in relatively lower overhead for the compressed representation in bfloat16 compared to int8.
This compressed format substantially reduces both on-chip and off-chip memory requirements, achieving up to 3.56$\times$ reduction (Table \ref{tb:sparsity_summary_benefits}).

Second, we notice that every sparsity level is
translated to its corresponding speedup, \emph{e.g.,} 4$\times$ for 1:4 (75\%) sparsity, achieving 100\% SPE utilization in all cases (Table \ref{tb:sparsity_summary_benefits}).
For all sparsity levels, data reuse is maximized since both indices and values are propagated and reused horizontally and vertically, similar to dense operation.

\subsection{GEMM Design Utilizing Multiple SST Slices}
\label{subsec:Matrix_multiplication_mapping}

In this section, we describe parametric GEMM implementations for both dense and sparse configurations, utilizing multiple SST slices.
Both implementations are highly regular and scale effectively on the FPGA fabric, attaining high frequencies as shown in Sec. \ref{subsec:Sparse_GEMM_implementation}.

\subsubsection{Dense Implementation}

Fig. \ref{fig:GEMM_2D_array_SSTs}a depicts a parametric GEMM accelerator comprising a 2D array of SST slices, which are configured in \textit{dense} mode (Sec. \ref{subsec:Sparse_Processing_Element}). 
The 2D array consists of $Y \cdot X$ SST slices, implementing a total SA size of $(Y \cdot 4) \times (X \cdot 4)$.
This size is denoted as the \textit{native} size of the GEMM accelerator. 
On-chip memory buffers are implemented to store the input matrices $A$, $B$ and the output matrix $C$.
The input buffers $A$, $B$ are located in the left and top edges of the 2D array, respectively, and are partitioned into banks, providing sufficient bandwidth to feed the SSTs.
In particular, for buffer $A$, $Y$ banks are required, while for buffer $B$, $X$ banks are needed.
Since each SST slice includes a 4$\times$4 SA, each bank needs to provide a bandwidth of 32-bits per cycle when SSTs are configured for int8 precision, while for bfloat16, 64-bits per cycle are required.
Regarding the output buffer $C$, $X \cdot Y$ banks are needed due to the output stationary architecture of the SSTs, each receiving an output of 128-bits per cycle (four 32-bit values as explained in Sec. \ref{subsec:Sparse_Tensor_slices_architecture}).

The data from the buffers $A$, $B$ are propagated in a systolic fashion between the SST slices, as illustrated in Fig. \ref{fig:GEMM_2D_array_SSTs}a.
Horizontally, the $A$ data are propagated via the global routing resources of the FPGA fabric.
Vertically, the $B$ data are inserted via global routing wires in the first SST at each vertical chain ($Y$ SSTs in Fig. \ref{fig:GEMM_2D_array_SSTs}a), while being forwarded to the next SSTs via dedicated wires.
It is important to note here that the SSTs comprising each vertical chain are \textit{physically} contiguous in the FPGA.
Nevertheless, in the horizontal dimension, the $Y$ chains might not follow the \textit{logical} arrangement shown in Fig. \ref{fig:GEMM_2D_array_SSTs}a inside the FPGA fabric, due to the routing flexibility of FPGAs.
This depends on decisions made by the FPGA place and route (PnR) algorithm.
Finally, notice the (static) systolic data setup configuration specifically for the SST slices that interface with the buffers $A$, $B$ (left and top edge of the 2D array).

Control logic, mapped to the CLB resources of the FPGA, is utilized to orchestrate the entire operation of the GEMM design.
We also implement tiling logic (in CLBs) to exploit data reuse in GEMM, as well as to support arbitrary GEMM sizes based on the available on-chip memory resources.
When mapping an arbitrary GEMM of $M^\prime$$\times$$K^\prime$$\times$$N^\prime$ dimensions, $M^\prime$ must be a multiple of $(Y \cdot 4)$, while $N^\prime$ must be a multiple of $(X \cdot 4)$, since the \textit{native} size of the accelerator is $(Y \cdot 4) \times (X \cdot 4)$.
Note that there is no constraint on the reduction $K^\prime$ dimension.
Finally, although not shown in Fig. \ref{fig:GEMM_2D_array_SSTs}a, $accumulate$ signals utilized during tiling (Sec. \ref{subsec:Sparse_Tensor_slices_architecture}) are propagated in a systolic fashion among the 2D array of SSTs, similar to the $A$, $B$ data.
The control logic sets the $accumulate$ signal only for the first SST in both vertical and horizontal dimensions (\emph{i.e.,} the SST fed by buffers $A_{1}$ and $B_{1}$), which significantly simplifies the overall logic.

\subsubsection{Dynamic Sparse Configuration}

Fig. \ref{fig:GEMM_2D_array_SSTs}b illustrates a parametric GEMM design that is \textit{dynamically} configured to support all the sparsity modes in the SSTs, \emph{i.e.,} dense, 2:4, 1:3 and 1:4.
This dynamic configuration is particularly important for layer-wise sparsity exploitation in DNNs, since each layer might require different sparsity level for optimal trade-off between DNN accuracy and speedup (Sec. \ref{subsec:Performance_estimation_DNNs}).
We note that the implementation is similar to the dense design (Fig. \ref{fig:GEMM_2D_array_SSTs}a), with main differences lying in the design of buffers $A$ and $B$, as well as in the vertical and horizontal propagation of the data.
Horizontally, the $A$ data are kept in the buffer banks in compressed format for the sparse modes (2:4, 1:3 and 1:4).
In this case, both non-zero data and indices are loaded in the SSTs and are propagated horizontally (see Sec. \ref{subsec:Sparse_Processing_Element}).
More specifically, for int8, each buffer $A$ bank needs to provide a bandwidth of 40-bits per cycle, due to the additional 8-bits indices for the four vertical SPEs at the interface of each SST.
Similarly, for bfloat16, 72-bits per cycle are required.

Vertically, four $B$ banks are needed to feed the SSTs due to the 4$\times$ increase in ports for 2:4 and 1:4 sparsity acceleration (Sec. \ref{subsec:Sparse_Processing_Element}).
Each $B$ bank provides the same bandwidth as the dense design (Fig. \ref{fig:GEMM_2D_array_SSTs}a), \emph{i.e.,} 32-bits and 64-bits per cycle for int8 and bfloat16, respectively.
Similar to the dense design, the $B$ data are inserted in the first SST at each vertical chain and dedicated wires are used to propagate them vertically (Fig. \ref{fig:GEMM_2D_array_SSTs}b).
These dedicated wires are particularly important for the sparse design, since otherwise 4$\times$ more vertical wires would be required to use global routing compared to the dense design (in the case where \textit{only} non-dedicated wires are employed).

Besides sparse operation, the design in Fig. \ref{fig:GEMM_2D_array_SSTs}b also supports dense computation.
This is because dense computation might still be needed, even if all weights in DNNs are sparse.
For instance, in Transformer-based DNNs \cite{Attention_all_you_need_2017, BERT_2019, ViT_2020}, the QKV (Query, Key, Value) 
GEMMs do not involve weights, and are typically computed as dense.
For dense computation, only one $B$ bank is sufficient at each vertical chain, \emph{e.g.,} $B_{x,0}$ (Fig. \ref{fig:GEMM_2D_array_SSTs}b).
However, multiplexing logic can be employed to 
utilize the remaining $B_{x,1}$, $B_{x,2}$, $B_{x,3}$ banks.
This is particularly important to ensure efficient utilization of on-chip memory resources, which leads to maximized data reuse and thus optimized energy efficiency \cite{Versal_vs_Stratix_FCCM_2024, MaxEVA_2023}.
Finally, for 1:3 sparsity, only banks $B_{x,0}$, $B_{x,1}$, $B_{x,2}$ are required, leaving bank $B_{x,3}$ unused. 
However, similar to the dense operation, 
multiplexing logic can be employed to reuse this bank when it comprises multiple BRAMs, which we do not explore in this work (see Sec. \ref{subsec:Sparse_GEMM_implementation} for implementation details).

The dynamic configuration among all sparsity levels 
is implemented in the control logic (using CLBs).
However, FPGA accelerators can be designed in a custom fashion depending on the sparsity of each DNN.
For instance, a specific DNN might require only 1:3 sparsity and dense computation across all of its layers.
The control and tiling logic for sparsity is similar to the dense design (Fig. \ref{fig:GEMM_2D_array_SSTs}a), showcasing a marginal increase in CLB resources (see Sec. \ref{subsec:Sparse_GEMM_implementation}).

\section{Evaluation}
\label{sec:Evaluation}

\begin{table}[t]
\vspace{-0.30cm}
\centering
\caption{FPGA logic tile and routing parameters in COFFE.}

\setlength\tabcolsep{1pt}
\resizebox{0.92\linewidth}{!}{
\begin{tabular}{cc|cc}
\Xhline{2.5\arrayrulewidth} 

\textbf{Parameter} & \textbf{Value} & \textbf{Parameter} & \textbf{Value} \\

\hline
\hline

FLEs per CLB ($N$) & 10 & Frac. LUT size ($K$) & 6\\
FLE independent inputs & 2 & Adders per FLE & 2\\
Channel width ($W$) & 300 & Wire length ($L$) & 4/16\\
CLB inputs ($I$) & 60 & CLB outputs ($O$) & 40\\
FLE outputs to routing ($O_{r}$) & 4 & Feedback FLE outputs ($O_{fb}$) & 2\\
SB flexibility ($F_{s}$) & 3 & Input connection flex. ($F_{cin}$) & 0.15\\
Output connection flex. ($F_{cout}$) & 0.1 & Input crossbar flex. ($F_{clocal}$) & 0.5\\

\Xhline{2.5\arrayrulewidth}

\end{tabular}
}

\label{tb:FPGA_tile_architectural_parameters}

\vspace{-0.50cm}

\end{table}

\subsection{FPGA Architecture}
\label{subsec:FPGA_architecture}

The traditional FPGA architecture comprises 
CLBs, BRAMs, DSP slices and routing resources, \emph{i.e.,} connection and switch boxes (CBs and SBs).
We enrich the FPGA architecture with our proposed in-fabric SST slices.
The SST columns repeat every 15 FPGAs columns, occupying only 14\% of the total FPGA area, which is sufficient for DNN applications as also found in \cite{Aman_TS_TRETS_2022} (refer to Fig. \ref{fig:GEMM_SST_3x3_int8} for the layout of the proposed FPGA architecture).
We exploit the automated transistor sizing tool, COFFE \cite{3D_FPGAs_COFFE_Boutros_FPT23, COFFE2_TRETS_2019}, to model the delays and areas of the FPGA components. 
These delays and areas are subsequently used to describe the FPGA architecture in the Verilog-to-Routing (VTR) tool flow \cite{VTR_8_2020}, facilitating FPGA architecture exploration.
We utilize the 7nm FinFET ASAP7 predictive process design kit (PDK) \cite{ASAP7_2016}, for the SPICE simulations conducted by COFFE.
Specifically, we exploit the typical corner (\emph{TT}) ASAP7 model \cite{ASAP7_2016, ASAP7_invited_2017}. 
Finally, COFFE was run with cost function of $area \times delay$ for four transistor sizing iterations (2--4 iterations are typically sufficient 
\cite{Original_COFFE_2013}).

We model a modern Agilex-like FPGA similar to \cite{TS_Aman_FPGA_2021, Aman_TS_TRETS_2022}.
Table \ref{tb:FPGA_tile_architectural_parameters} shows the FPGA logic (CLB) and routing (CBs and SBs) architectural parameters used in COFFE.
Each CLB contains 10 fracturable logic elements (FLEs), which consist of 6-input lookup tables (LUTs), registers and two adder bits.
Since COFFE does not model 7nm optimized memory circuits, we obtain the model of the 20Kbit BRAMs used in the recent work \cite{3D_FPGAs_COFFE_Boutros_FPT23} for 7nm FPGAs.
The delay and area values of these BRAMs are conservatively estimated using the 14nm Stratix 10 architecture values, as described in \cite{3D_FPGAs_COFFE_Boutros_FPT23}. 
Finally, we leverage the DSP used in \cite{Aman_TS_TRETS_2022}, which supports multiple precisions and modes, closely matching the commercial Agilex DSP \cite{Intel_Agilex_DSP_2020}.
However, since this DSP is modeled in 22nm, we scale down its area and delay values to 7nm, exploiting the scaling factors from \cite{Scale_VLSI_2017}.

\subsection{SST Slices Implementation}
\label{subsec:SST_slices_implementation}

We implement the SST slices utilizing the  ASAP7 7.5--track standard cell library \cite{ASAP7_invited_2017, ASAP7_github}.
Specifically, we use the regular threshold voltage (RVT) cells of the typical (\emph{TT}) corner of the library.
COFFE's hybrid flow was exploited, where the core of the SSTs is implemented using the standard cell flow, while the interface to the programmable routing (local crossbar, drivers for dedicated wires, etc.) is implemented in the full custom flow using SPICE simulations.
The standard cell flow uses Synopsys Design Compiler for synthesis, Cadence Innovus for PnR and Synopsys Primetime for timing analysis.
The full custom flow uses the \emph{TT} ASAP7 SPICE model and COFFE was run with the same configurations as in the previous section.

\begin{table}[t]
\vspace{-0.30cm}
\centering
\caption{Area and freq. of SST \emph{vs.} SDT\_GIO slices (post PnR).}

\setlength\tabcolsep{2.5pt}
\renewcommand{\arraystretch}{1.0}
\resizebox{0.80\linewidth}{!}{
\begin{tabular}{c|ccc}
\Xhline{2.5\arrayrulewidth}

\multicolumn{2}{c}{\textbf{In-fabric slice}} & \textbf{SST} & \textbf{SDT\_GIO}\\
\hline
\hline
& Standard-cell core & 4530.8 (+22.9\%) & 3687.8 \\

& Input crossbar & 1511.4 & 985.6\\

\textbf{Area} & Dedicated output routing & 40.2 & 0\\

\textbf{($\boldsymbol{\mu m^2}$)} & Switch Box (SB) & 1374.3 & 1603.4\\

& Connection Box (CB) & 734.9 & 571.6\\

& \textbf{Total tile} & \textbf{8191.6 (+19.6\%)} & \textbf{6848.4}\\

\hline

\textbf{Freq.} & 
int8 & 928.6 & 935.3 \\

\textbf{(MHz)} & bfloat16 & 838.1 & 850.5\\

\Xhline{2.5\arrayrulewidth}

\end{tabular}
}

\label{tb:SST_delay_area_vs_SDT_GIO}

\vspace{-0.55cm}

\end{table}

To accurately quantify the area overhead of the sparsity enhancements described in Sec. \ref{subsec:Sparse_Processing_Element}, we remove the sparsity features of the SST slices, to implement a 2D systolic dense tensor slice.
We retain all the design features delineated in Sec. \ref{subsec:Sparse_Tensor_slices_architecture}, except the dedicated wires.
For this dense tensor slice, we utilize the approach proposed in \cite{TS_Aman_FPGA_2021, Aman_TS_TRETS_2022}, where all I/O ports are connected to global routing.
We refer to this tensor slice as SDT\_GIO, which is similar to \cite{TS_Aman_FPGA_2021, Aman_TS_TRETS_2022}.
The dense SDT\_GIO slices are used as the baseline comparison with our proposed sparse SSTs, while also allowing for quantification of the routing savings due to dedicated wires in SSTs 
(Sec. \ref{subsec:Dedicated_wires_benefits}).

Table \ref{tb:SST_delay_area_vs_SDT_GIO} shows the area and frequency results for both the SST and SDT\_GIO slices obtained from COFFE. 
At the standard-cell core level, we observe a low area overhead of 22.9\% for the sparsity enhancements of the SST slices.
For both SST and SDT\_GIO slices, we implement a 50\% populated input crossbar to enhance their routability inside the FPGA fabric \cite{COFFE2_TRETS_2019}. 
Each SST slice has 333 global routing inputs and 202 outputs, along with 256 dedicated inputs/outputs.
In contrast, the SDT\_GIO slice has 259 global routing inputs and 258 outputs, without dedicated I/Os.
Given the aforementioned global routing I/Os, the SST slices need to access 6 SBs, thus spanning 6  CLB (logic) tiles.
However, since more outputs are needed for the SDT\_GIOs (primarily due to absence of dedicated wires), they need to span 7 CLB tiles.
Therefore, when calculating the total tile area for both slices (standard-cell core and routing interface), we obtain a low area increase of \textbf{19.6\%} for 
SSTs \emph{vs.} SDT\_GIOs (Table \ref{tb:SST_delay_area_vs_SDT_GIO}).
Finally, we observe that both slices can operate at high frequencies for the supported precisions (> 838 MHz in all cases), similar to commercial 7nm FPGA hard blocks \cite{Versal_DSP_frequencies, Agilex_5_frequencies}.



\begin{figure}[t]
\vspace{-0.50cm}
\centering
\subfloat[int8]{\includegraphics[width=0.48\linewidth]{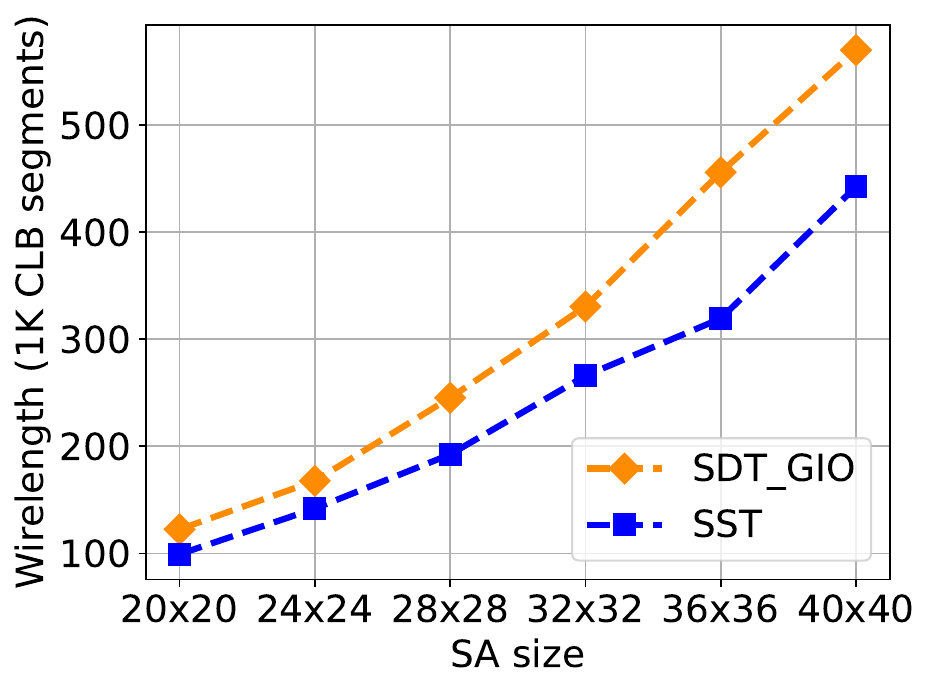}
\label{fig:wirelength_SST_vs_GIO_int8}}
\hfill
\subfloat[bfloat16]{\includegraphics[width=0.48\linewidth]{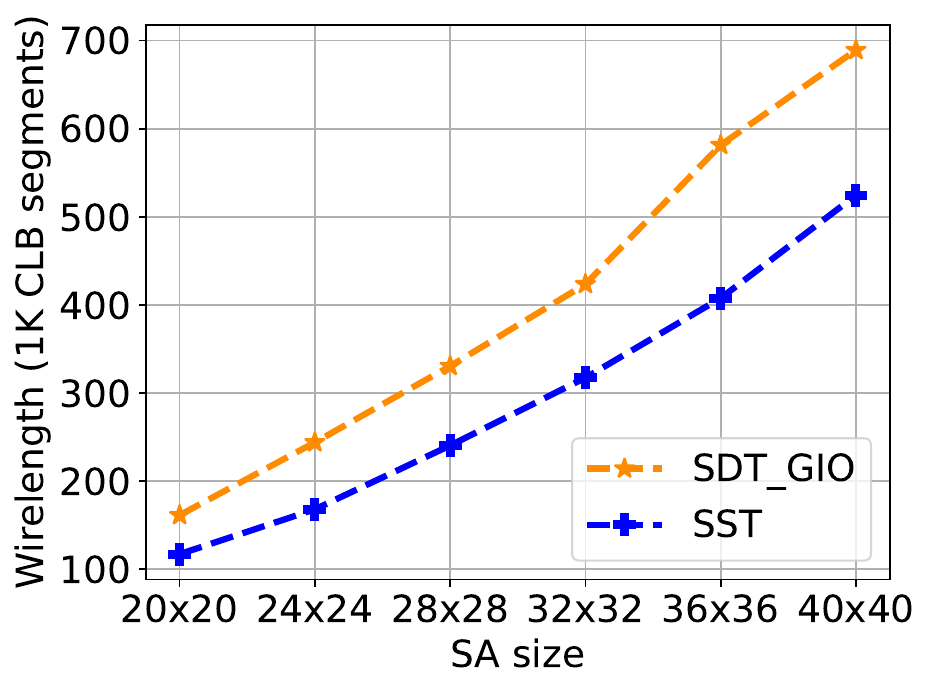}
\label{fig:wirelength_SST_vs_GIO_bf16}}

\vspace{-0.3cm}

\caption{Routing wirelength comparison of \textit{dense} GEMM mapped to SST and SDT\_GIO slices, for various SA sizes.}
\label{fig:wirelength_SST_vs_GIO}
\vspace{-0.45cm}
\end{figure}

\subsection{Wirelength Gains Utilizing Dedicated Wires}
\label{subsec:Dedicated_wires_benefits}

To quantify the benefits of the proposed dedicated wires in SSTs, we implement \textit{dense} GEMMs utilizing our SSTs slices (Fig. \ref{fig:GEMM_2D_array_SSTs}a) and compare them with \textit{dense} GEMMs using SDT\_GIOs slices, which support only global routing.
To this end, similar to the enhanced FPGA architecture with SST slices, we create an architecture that has SDT\_GIOs instead of SSTs. 
To facilitate the exploration, we implement a parametric Python script that generates the Verilog code for arbitrary GEMM sizes.
The Verilog designs are synthesized and implemented on the two aforementioned FPGA architectures (one with SST and one with SDT\_GIO slices), exploiting the VTR flow.
In all cases, we perform a 10 seed-sweep in VTR, and report the design that attains the maximum FPGA frequency.



Fig. \ref{fig:wirelength_SST_vs_GIO} illustrates the total routing wirelength of the two dense GEMM accelerators, for various SA sizes at both int8 and bfloat16 precisions.
These designs use 512-element deep banks for buffers $A$, $B$ and $C$ (Sec. \ref{subsec:Matrix_multiplication_mapping}), exploiting the 512x40-bit BRAM configuration mode (also found in recent Intel FPGAs \cite{Intel_Agilex_5_M20Ks, Intel_Agilex_7_M20Ks, Intel_Stratix_10_M20Ks}).
For the SA sizes shown in Fig. \ref{fig:wirelength_SST_vs_GIO}, we observe a significant wirelength reduction due to dedicated wires in the SSTs, ranging from \textbf{15.5--29.9\%} for int8 precision, compared to the SDT\_GIO-based GEMM designs.
Similarly, for bfloat16, the wirelength reduction ranges from \textbf{23.9--31.2\%}, showcasing the importance of vertical dedicated wires for 2D GEMM implementations on FPGA architectures.

Regarding the maximum attainable FPGA frequency, both dense GEMM implementations 
achieve high frequencies ranging from 668--731 MHz.
In particular, the smallest designs, \emph{i.e.,} 20x20 SA size, attain the highest frequency of $\sim$731 MHz for both SSTs and SDT\_GIOs at int8 and bfloat16 precisions.
Similarly, the largest designs, \emph{i.e.,} 40x40 SA size, 
both attain $\sim$668 MHz. 
We notice that dense GEMMs mapped to SST and SDT\_GIO slices attain almost similar FPGA frequencies (<1\% difference for all tested SA sizes depicted in Fig. \ref{fig:wirelength_SST_vs_GIO}).
This can be explained by the fact that the critical path is typically in the control logic implemented in CLBs, which is the same for both dense GEMMs mapped to SSTs and SDT\_GIOs. 


\subsection{Sparse SST-based GEMM Implementation}
\label{subsec:Sparse_GEMM_implementation}

In this section, we present the \textit{sparse} GEMM implementation shown in Fig. \ref{fig:GEMM_2D_array_SSTs}b, which allows dynamic configuration for all supported sparsity modes in SSTs, \emph{i.e.,} dense, 2:4, 1:3 and 1:4.
We use two baselines for our comparison.
First, we compare with a GEMM design mapped to SDT\_GIOs using a FPGA architecture that replaces SSTs with SDT\_GIOs. 
Since SDT\_GIOs have a 2D dense systolic dataflow that does not support sparse computation (Sec. \ref{subsec:SST_slices_implementation}), 
we retain the zeros of the sparse matrices, therefore operating similarly to dense. 
Hence, this implementation is similar to that shown in the previous section (Sec. \ref{subsec:Dedicated_wires_benefits}), however it has the same amount of on-chip memory as the sparse SST-based design. 
In particular, we implement four memory banks in the vertical dimension to feed each SDT\_GIO (similar to Fig. \ref{fig:GEMM_2D_array_SSTs}b).
This allows for the same data reuse opportunities as in the SST-based design, thus enabling a fair comparison.
Second, we implement the same sparse SST-based design, mapped to CLBs and DSPs of a traditional FPGA architecture, \emph{i.e.,} without SSTs and SDT\_GIOs.
Finally, we note that all tested designs utilize 512-element deep banks and are based on a 10 seed-sweep maximization of FPGA frequency, similar to Sec. \ref{subsec:Dedicated_wires_benefits}.

\subsubsection{FPGA Frequency}
\begin{figure}[t]
\vspace{-0.75cm}
\centering
\subfloat[int8]{\includegraphics[width=0.48\linewidth]{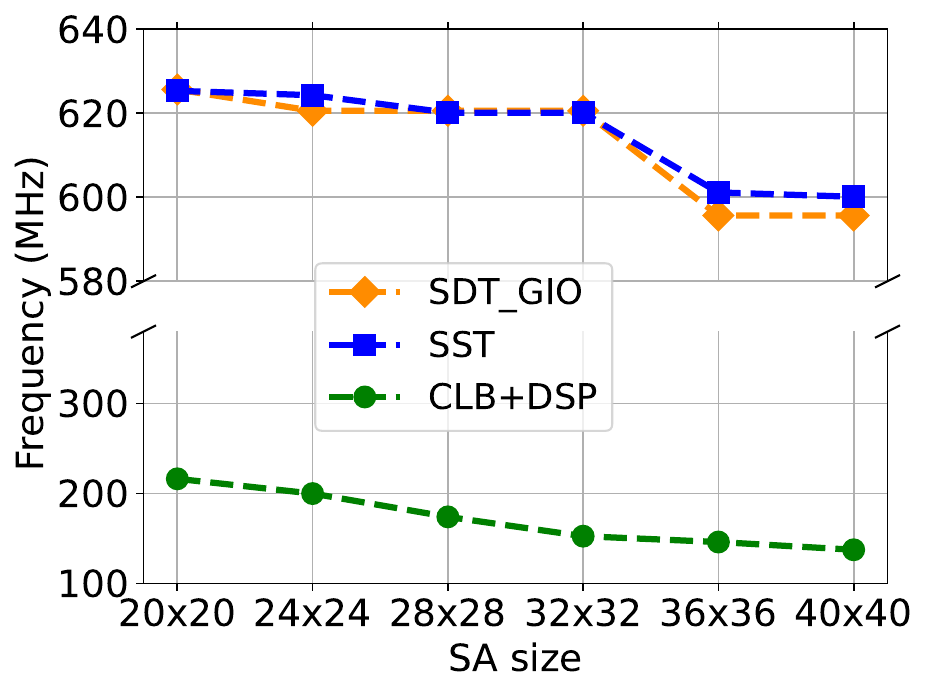}
\label{fig:freq_SST_GIO_DSP_int8}}
\hfill
\subfloat[bfloat16]{\includegraphics[width=0.48\linewidth]{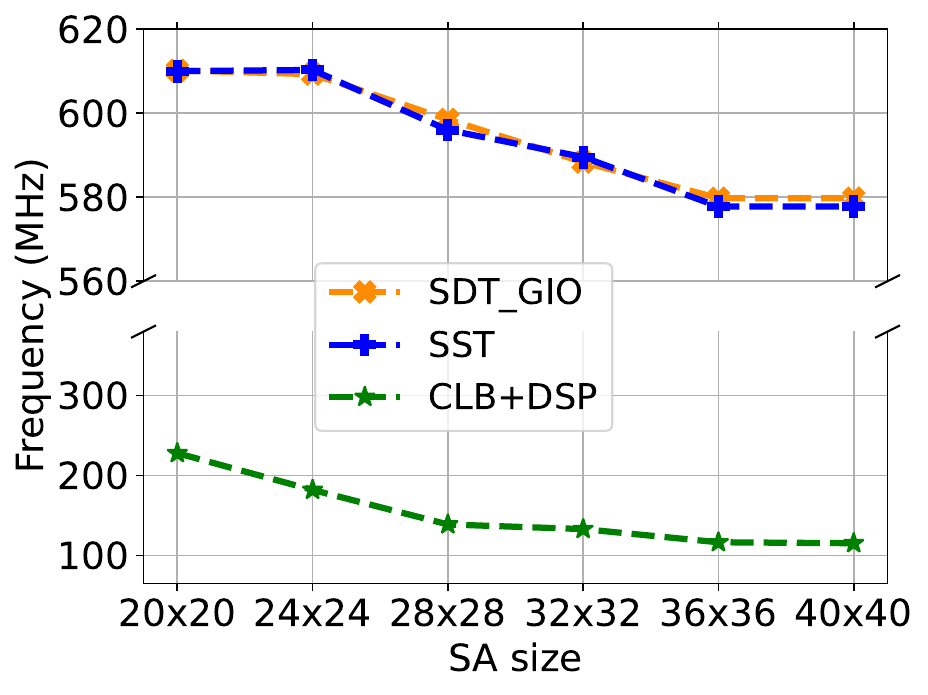}}
\label{fig:freq_SST_GIO_DSP_bf16}

\vspace{-0.3cm}

\caption{FPGA frequency comparison of \textit{sparse} GEMM using SSTs \emph{vs.} SDT\_GIOs \emph{vs.} CLBs+DSPs, for various SA sizes.}
\label{fig:freq_SST_vs_GIO_vs_DSP}
\vspace{-0.55cm}
\end{figure}

Fig. \ref{fig:freq_SST_vs_GIO_vs_DSP} depicts the FPGA frequency of the sparse GEMMs mapped to SST slices, SDT\_GIO slices as well as CLBs and DSPs.
First, we observe the high FPGA frequency of GEMMs mapped to SST and SDT\_GIO slices, ranging from 596--625 MHz for int8 and from 578--610 MHz for bfloat16. 
We note that the slightly lower frequencies of the SDT\_GIO-based GEMMs compared to the previous section are attributed to the higher BRAM usage.
For all tested SA sizes, we notice a negligible difference in FPGA frequency (<1\%) between the SST-based and the SDT\_GIO-based GEMMs. 
However, when comparing the SST-based \emph{vs.} the CLB+DSP GEMMs, we observe considerably higher frequencies, ranging from \textbf{2.8--4.4$\times$} and from \textbf{2.7--5$\times$} for int8 and bfloat16, respectively.

Second, note that the sparse GEMMs utilizing SSTs scale effectively when increasing the SA size.
In particular, we observe a minor decrease in FPGA frequency (4.6\% and 5.2\% for int8 and bfloat16, respectively), when considering the 40x40 \emph{vs.} the 20x20 SA size.
In contrast, the sparse GEMM mapped to CLBs and DSPs, does not scale as effectively, since we notice a substantial frequency drop of 36.6\% and 49.6\% for int8 and bfloat16, respectively.
These results emphasize the significant performance advantages and scalability of in-fabric sparse hard blocks compared to traditional FPGAs.





\subsubsection{GEMM Area}

\begin{figure}[t]

\vspace{-0.25cm}

\centering
\subfloat[int8]{\includegraphics[width=0.48\linewidth]{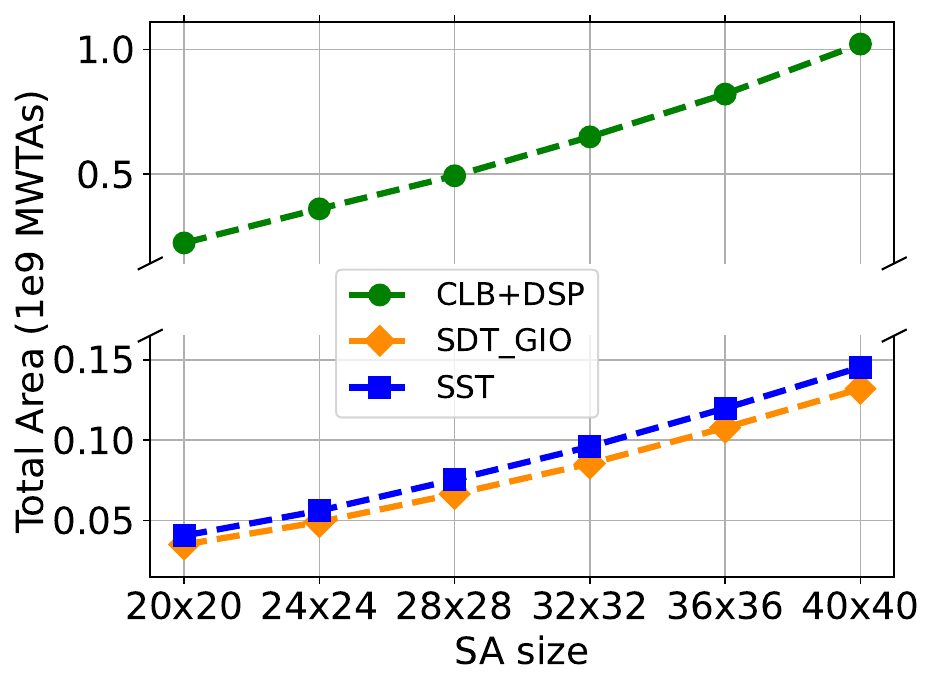}}
\label{fig:area_SST_GIO_DSP_int8}
\hfill
\subfloat[bfloat16]{\includegraphics[width=0.48\linewidth]{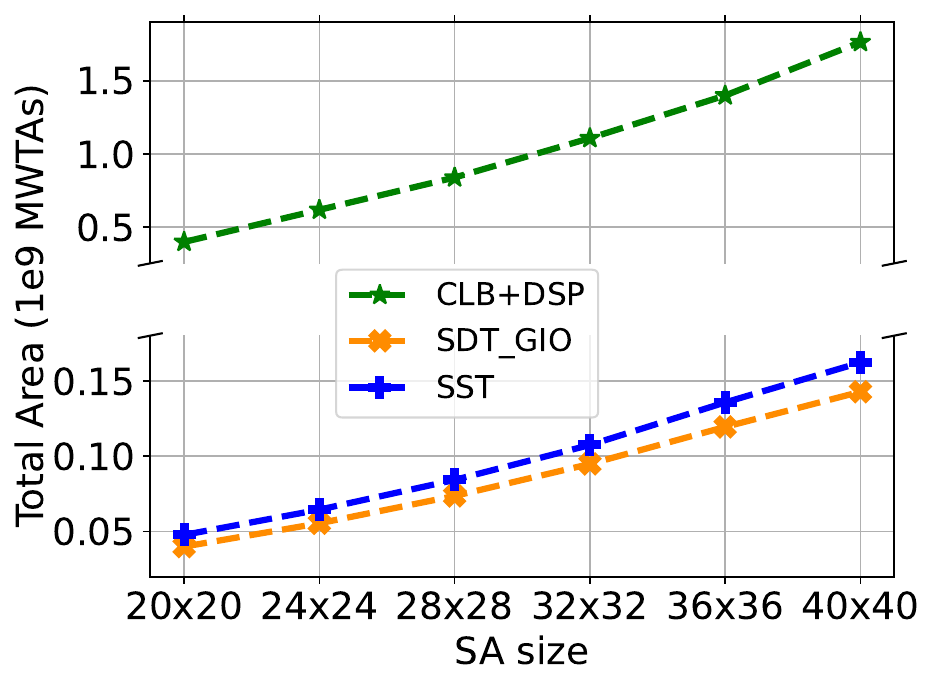}}
\label{fig:area_SST_GIO_DSP_bf16}

\vspace{-0.3cm}

\caption{Total area comparison of various \textit{sparse} GEMM implementations utilizing SSTs \emph{vs.} SDT\_GIOs \emph{vs.} CLBs+DSPs.}
\label{fig:area_SST_vs_GIO_vs_DSP}

\vspace{-0.55cm}

\end{figure}

We calculate the total area of our GEMM designs as the summation of the utilized logic area and the FPGA routing area.
While the utilized logic area is reported in VTR, the tool  does not report the used routing area.
Thus, we estimate the routing area by summing the area of all utilized multiplexers in both the SB and CB resources of our designs.
Fig. \ref{fig:area_SST_vs_GIO_vs_DSP} shows the total area of our sparse GEMM designs in minimum width transistor area (MWTA) units \cite{VTR_8_2020}.
As shown in Table \ref{tb:SST_delay_area_vs_SDT_GIO}, we observed a 19.6\% area overhead of the SSTs \emph{vs.} the SDT\_GIOs at the \textit{tile level}.
However, here we calculate this area overhead at the \textit{GEMM implementation level}, as this provides more accurate estimations for DNN accelerators.
When comparing the sparse GEMM mapped to SSTs \emph{vs.} the GEMM using SDT\_GIOs, we observe a small area increase ranging from \textbf{10.2--15.9\%} for int8, and from \textbf{13.3--19.4\%} for bfloat16, across all SA sizes.
We note that the lowest area increase, \emph{i.e.,} \textbf{10.2\%} and \textbf{13.3\%} for int8 and bfloat16, respectively, occurs for the highest, 40x40 SA size.
This is mainly attributed to the dedicated wires in SSTs, where the increase of the SA size (thus the total slices), leads to more routing area in the GEMM design using SDT\_GIOs. 

As shown in Fig. \ref{fig:area_SST_vs_GIO_vs_DSP}, we notice a substantial area increase of the sparse GEMM mapped to CLBs and DSPs \emph{vs.} using SSTs, of up to \textbf{7$\times$} for int8, and \textbf{10.9$\times$} for bfloat16.
Moreover, we observe that this area difference increases with the SA size.
In particular, for 20x20 SA size, we obtain a difference of 5.5$\times$ and 8.3$\times$ for int8 and bfloat16, respectively.
However, for 40x40 size, the difference increases to 7$\times$ and 10.9$\times$ for int8 and bfloat16, respectively.
Hence, our results exhibit the significant area efficiency and scalability of the proposed sparse GEMM designs using in-fabric SSTs, 
over traditional FPGAs.








\subsubsection{Resource Usage \& Efficiency of Sparse SST-based GEMM}

Table \ref{tb:GEMM_SST_vs_SDT_GIO_vs_DSP_40x40_SA} presents the resource usage and several evaluation metrics of the sparse GEMM designs, for SA size of 40x40.
First, we notice a small CLB increase for the SST \emph{vs.} the SDT\_GIO GEMM, of 26.2\% and 29.1\% for int8 and bfloat16, respectively.
This is ascribed to the additional control logic of including all supported sparsity levels, compared to only control logic for dense (since for SDT\_GIOs the zeros of the sparse matrices are retained, effectively operating similar to dense).
Second, note that the BRAM usage is the same in all GEMM designs, despite the additional bandwidth requirement due to the indices of the compressed sparse format (Sec. \ref{subsec:Matrix_multiplication_mapping}).
This is due to the use of the 512x40-bit BRAM mode (Sec. \ref{subsec:Dedicated_wires_benefits}). 
For example, for int8, buffer $A$ banks in the dense design of Fig. \ref{fig:GEMM_2D_array_SSTs}a, need a bitwidth of 4$\cdot$8=32-bits, which is smaller than the 40-bits width of the BRAMs.
However, for the sparse operation of Fig. \ref{fig:GEMM_2D_array_SSTs}b, they need additionally 8-bits for the indices (40-bits total), which is exactly the same as the BRAMs bitwidth.
Similar conclusions can be drawn for bfloat16.
Hence, no additional BRAMs are needed for the higher bandwidth requirements of sparse matrices in compressed format.

Table \ref{tb:GEMM_SST_vs_SDT_GIO_vs_DSP_40x40_SA} shows the \textit{effective} throughput and area efficiency of our GEMM designs.
The \textit{effective} throughput (TOPs) is calculated on the GEMM \textit{native} size (40x40 in this case, see Sec. \ref{subsec:Matrix_multiplication_mapping}), and similar to \cite{Nvidia_accelerate_sparse_2021, NVIDIA_A100, AMD_CDNA_3} when acceleration can be attained from sparsity.
In particular, since 1:4 sparsity is included in our SST-based and CLB+DSP designs, their \textit{effective} throughput is increased by a factor of four.
When comparing the SST-based GEMM \emph{vs.} the SDT\_GIO-based design, we observe a throughput gain of \textbf{4.03$\times$} and \textbf{3.98$\times$} for int8 and bfloat16, respectively.
Moreover, we observe \textbf{3.66$\times$} (int8) and \textbf{3.51$\times$} (bfloat16) higher area efficiency (TOPs/GMWTA, area in Giga MWTAs).
In addition, when comparing with the GEMM using CLBs and DSPs, we obtain \textbf{4.34$\times$} (int8) and \textbf{5$\times$} (bfloat16) higher throughput for the SST-based GEMM.
Regarding the area efficiency, an immense \textbf{30.62$\times$} (int8) and \textbf{54.45$\times$} (bfloat16) difference is observed, primarily due to the increased usage
of the traditional FPGA resources.
To this end, notice the very high CLB and wirelength usage for the CLB+DSP \emph{vs.} the SST-based GEMM. 
For instance, for int8, there is a 34.1$\times$ higher CLB and a 15$\times$ higher wirelength usage.

\begin{figure}[t]
\vspace{-0.30cm}
\centering
\includegraphics[width=0.52\linewidth]{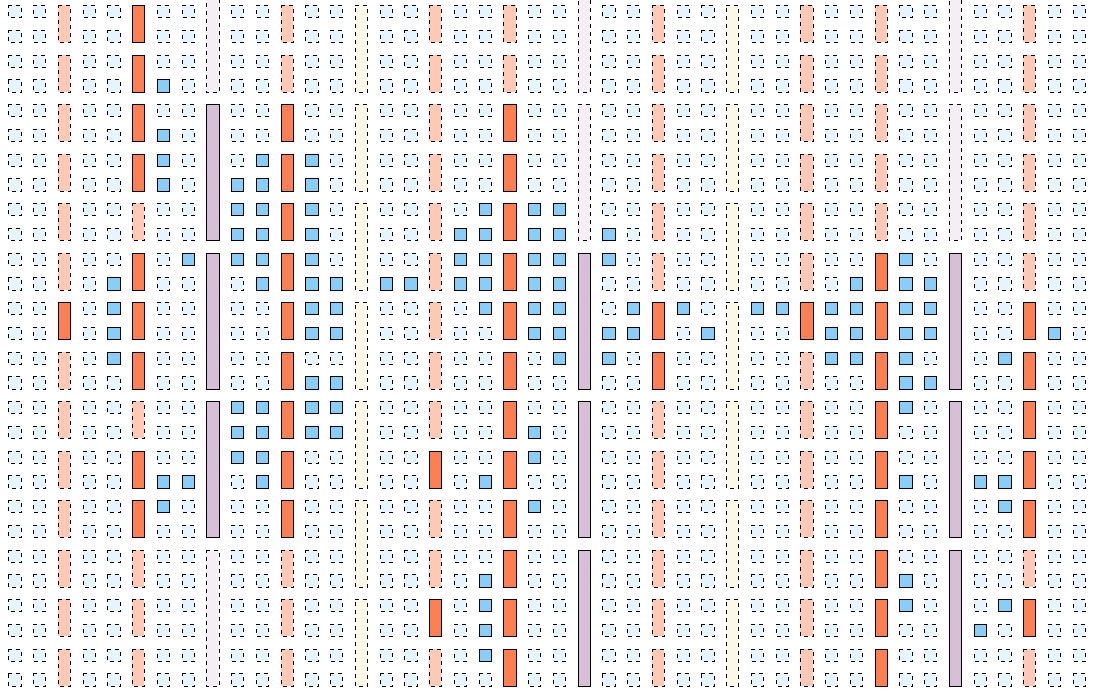}

\vspace{-0.3cm}

\caption{SST-based GEMM implemented in VTR (SA size: 12x12). Blue: CLB, Orange: BRAM, Yellow: DSP, Purple: SST.}
\label{fig:GEMM_SST_3x3_int8}

\vspace{-0.50cm}

\end{figure}

Finally, in Fig. \ref{fig:GEMM_SST_3x3_int8} we show a screenshot of a sparse SST-based design implemented in VTR.
Notice the physically contiguous vertical SST chains due to dedicated wires.
Also, observe the 2D \textit{physical} layout after VTR PnR, compared to the \textit{logical} 2D array of Fig. \ref{fig:GEMM_2D_array_SSTs}b.

\begin{table*}[t]
\vspace{-0.40cm}
 \centering
\caption{Evaluation and comparison of sparse GEMM designs using SSTs \emph{vs.} SDT\_GIOs \emph{vs.} CLBs+DSPs (SA size: 40x40).}
\vspace{-0.10cm}
\setlength\tabcolsep{2.8pt}
\renewcommand{\arraystretch}{1.15}
\resizebox{0.81\textwidth}{!}{
\vspace{-0.10cm}
\begin{tabular}{c|c|c|c|c|c|c|c|c|c|c|c}
\Xhline{2.5\arrayrulewidth}


\multirow{2}{*}{\textbf{Pr.}} & \textbf{GEMM} & \multirow{2}{*}{\textbf{CLBs}} & \multirow{2}{*}{\textbf{BRAMs}} & \textbf{DSP} & \textbf{SST} & \textbf{SDT\_GIO} & \textbf{Freq.} & \textbf{Area} & \textbf{Wirelength} & \textbf{Eff. Thrpt.} & \textbf{Area Eff.}\\

& \textbf{Cfg.} &  &  & \textbf{slices} & \textbf{slices} & \textbf{slices} & \textbf{(MHz)} & \textbf{(GMWTA)} & \textbf{(CLB seg. units)} & \textbf{(TOPs)} & \textbf{(TOPs/GMWTA)}\\

\hline
\hline

\multirow{3}{*}{\rotatebox[]{90}{\textbf{int8}}} & SST  & 665 & 450 & 0 & 100 & -- & 601 &  0.145 & 493K & \textbf{7.69} & \textbf{53.03}\\

& SDT\_GIO & 527 & 450 & 0 & -- & 100 & 596 & 0.132 & 578K & 1.91 & 14.47\\

& CLB+DSP & 22650 & 450 & 400 & -- & -- & 138 & 1.022 & 7373K & 1.77 & 1.732 \\

\hline

\multirow{3}{*}{\rotatebox[]{90}{\textbf{bfloat16}}} & SST & 719 & 500 & 0 & 100 & -- & 578 &  0.162 & 684K & \textbf{7.40} & \textbf{45.68}\\

& SDT\_GIO & 557 & 500 & 0 & -- & 100 & 580 & 0.143 & 725K & 1.86 & 13.01\\

& CLB+DSP & 27065 & 500 & 2400 & -- & -- & 116 & 1.764 & 15212K & 1.48 & 0.839 \\

\Xhline{2.5\arrayrulewidth}

\end{tabular}
}
\label{tb:GEMM_SST_vs_SDT_GIO_vs_DSP_40x40_SA}

\vspace{-0.20cm}

\end{table*}

\begin{table*}[t]


\centering
\caption{Speedup estimation, accuracy and weight mem. reduction of various DNNs when mapped to SSTs \emph{vs.} SDT\_GIOs.}
\vspace{-0.10cm}
\setlength\tabcolsep{2.5pt}
\renewcommand{\arraystretch}{1.0}
\resizebox{0.81\textwidth}{!}{
\vspace{-0.10cm}
\begin{tabular}{c|c|c|c|c|c|c|c|c|c||c|c}
\Xhline{2.5\arrayrulewidth}


\textbf{DNN} & \textbf{Sparsity} & \multicolumn{4}{c|}{\textbf{Number of layers}} & \textbf{In-fabric} & \multirow{2}{*}{\textbf{Speedup}} & \multicolumn{2}{c||}{\textbf{Top-1 accuracy (\%)}} & \multicolumn{2}{c}{\textbf{Weight reduction}} \\

\cline{3-6} 
\cline{9-12} 

\textbf{model} & \textbf{levels} & \textbf{Dense} & \textbf{2:4} & \textbf{1:3} & \textbf{1:4} & \textbf{slice} & & \textbf{int8} & \textbf{bfloat16} & \textbf{int8} & \textbf{bfloat16} \\

\hline
\hline
 
 & dense & 48 & -- & -- & -- & SDT\_GIO & 1$\times$ & 79.56 (-0.00\%) & 79.79 (-0.00\%) & 1$\times$ & 1$\times$\\

\textbf{DeiT-S} & \textbf{[dense, 2:4]} & \textbf{--} & \textbf{48} & \textbf{--} & \textbf{--} & \textbf{SST} & \textbf{1.88$\times$} & \textbf{79.08 (-0.48\%)} & \textbf{79.20 (-0.59\%)} & \textbf{1.58$\times$} & \textbf{1.76$\times$} \\

\textbf{(4.7 GFLOPs)} & [dense, 2:4, 1:3, 1:4] & 3 & 26 & 16 & 3 & SST & 2.14$\times$ & 77.95 (-1.61\%) & 78.33 (-1.46\%) & 1.81$\times$ & 2.02$\times$ \\

& [dense, 1:3] & -- & -- & 48 & -- & SST & 2.61$\times$ & 73.36 (-6.20\%) & 73.95 (-5.84\%) & 2.32$\times$ & 2.61$\times$ \\

\hline

 & dense & 48 & -- & -- & -- &  SDT\_GIO & 1$\times$ & 81.40 (-0.00\%) & 81.80 (-0.00\%) & 1$\times$ & 1$\times$\\

\textbf{DeiT-B} & [dense, 2:4] & -- & 48 & -- & -- & SST & 1.91$\times$ & 81.45 (+0.05\%) & 81.73 (-0.07\%) & 1.59$\times$ & 1.77$\times$ \\

\textbf{(17.6 GFLOPs)} & [dense, 2:4, 1:3, 1:4] & 6 & 15 & 13 & 14 & SST & 2.35$\times$ & 81.28 (-0.12\%) & 81.60 (-0.20\%) & 2.05$\times$ & 2.26$\times$ \\

& [dense, 1:3] & -- & -- & 48 & -- & SST & 2.75$\times$ & 81.10 (-0.30\%) & 81.25 (-0.55\%) & 2.36$\times$ & 2.64$\times$ \\

& \textbf{[dense, 1:4]} & \textbf{--} & \textbf{--} & \textbf{--} & \textbf{48} & \textbf{SST} & \textbf{3.52$\times$} & \textbf{80.37 (-1.03\%)} & \textbf{80.69 (-1.11\%)} & \textbf{3.12$\times$} & \textbf{3.50$\times$} \\

\hline

& dense & 108 & -- & -- & -- & SDT\_GIO & 1$\times$ & 82.99 (-0.00\%) & 83.09 (-0.00\%) & 1$\times$ & 1$\times$\\

\textbf{ConvNeXt-S}& [dense, 2:4] & 36 & 72 & -- & -- & SST & 1.93$\times$ & 82.52 (-0.47\%) & 82.62 (-0.47\%) & 1.46$\times$ & 1.66$\times$ \\

\textbf{(8.7 GFLOPs)}& \textbf{[dense, 2:4, 1:3, 1:4]} & \textbf{36} & \textbf{59} & \textbf{7} & \textbf{6} & \textbf{SST} & \textbf{2.07$\times$} & \textbf{82.30 (-0.69\%)} & \textbf{82.39 (-0.70\%)} & \textbf{1.52$\times$} & \textbf{1.74$\times$} \\

& [dense, 1:3] & 36 & -- & 72 & -- & SST &  2.81$\times$ & 81.44 (-1.55\%) & 81.53 (-1.56\%) & 1.95$\times$ & 2.32$\times$ \\

& [dense, 1:4] & 36 & -- & -- & 72 & SST & 3.63$\times$ & 81.05 (-1.94\%) & 81.16 (-1.93\%) & 2.34$\times$ & 2.89$\times$ \\

\Xhline{2.5\arrayrulewidth}

\end{tabular}
}
\label{tb:DNN_sparsity_results}

\vspace{-0.40cm}

\end{table*}


\subsection{DNN Speedup Estimation Using SST Slices}
\label{subsec:Performance_estimation_DNNs}

We provide a speedup estimation for actual DNNs when leveraging our proposed SST slices.
In particular, we utilize the DeiT \cite{DeiT_2021} and ConvNeXt \cite{Convnext_2022} models, which attain state-of-the-art accuracy for vision tasks.
We apply two types of sparsity to determine the optimal sparsity configuration for each aforementioned models. 
First, we utilize \textit{uniform} sparsity, \emph{i.e.,} all layers have the same sparsity, by pruning weights based on their magnitudes \cite{Nvidia_accelerate_sparse_2021}.
Second, we exploit \textit{layer-wise} sparsity by adopting the neural architecture search (NAS) methodology proposed in \cite{Huang_2024_CVPR} to identify the optimal sparsity level of each layer, aiming to minimize accuracy degradation. 
Accuracy is evaluated on the ImageNet-1K dataset \cite{deng2009imagenet}, with int8 quantization applied using the post-training method proposed in \cite{PTQ4ViT_arixv2022}.

For speedup estimation, we develop an analytical model based on our results in Sec. \ref{subsec:Sparse_GEMM_implementation}, comparing the sparse SST-based GEMM to the SDT\_GIO-based GEMM, with the latter operating on dense matrices.
Our model exploits the matrix sizes for each layer of the DeiT and ConvNeXt models, and applies zero padding to align with the \textit{native} size of the GEMM accelerator (Sec. \ref{subsec:Matrix_multiplication_mapping}).
We utilize our largest implemented GEMM design, \emph{i.e.,} 40x40 SA size, and assume 100 GB/s bandwidth for DRAM modeling, which is typical in modern FPGAs \cite{Speedster_product_brief, VCK_5000}.
Utilizing GEMM computations provides a reliable model for estimating speedup, since GEMM is the core computation in contemporary DNNs, accounting for more than 90\% of the total execution time \cite{adolf2016fathom, wang_gemm_breakdown}.
Other operations \emph{e.g.,} softmax, layernorm, can be effectively overlapped by DNN accelerators.
Moreover, GEMM-based estimation (convolutions implemented as GEMM for ConvNeXt) is also performed in multiple works targeting speedup calculations due to sparsity exploitation \cite{HighLight_MIT_2023, Vegeta_HPCA_2023, Sparse_tensor_GPUs_2019, STA_arxiv_2020, SDP_sparse_2023}.

\vspace{2mm}
 
Table \ref{tb:DNN_sparsity_results} presents the speedup estimation and accuracy for our two ViT models, \emph{i.e.,} DeiT-S and DeiT-B, and the ConvNeXt-S model. 
When applying uniform 2:4 sparsity, we observe a negligible accuracy decrease in DeiT-B for bfloat16 (0.07\%), or even slightly higher accuracy for int8 (by +0.05\%).
However, we notice a larger decrease for DeiT-S, \emph{e.g.,} 0.48\% for int8.
This is due to the fact that smaller models (DeiT-S with 4.7 GFLOPs) exhibit less redundancy compared to larger models (DeiT-B with 17.6 GFLOPs), making them more resilient to sparsity.
Note that even with uniform 2:4 sparsity across all layers, dense computation is still needed (Table \ref{tb:DNN_sparsity_results}), due to QKV computation in ViTs (Sec. \ref{subsec:Matrix_multiplication_mapping}).
In addition, for ConvNeXt-S, we retain its depth-wise convolution layers as dense, since we observed a severe accuracy degradation when attempting to sparsify them.
For instance, when applying 2:4 sparsity to all other layers except the (36) depth-wise layers in ConvNeXt-S, we obtain an accuracy degradation of only 0.47\%.
However, uniform 2:4 sparsity can result in significant speedup, \emph{e.g.,} 1.93$\times$ in ConvNeXt-S, 
compared to dense operation.
We note that both int8 and bfloat16 cases deliver nearly indistinguishable speedups over dense computation, since the frequency difference between the SST-based and SDT\_GIO-based GEMM accelerators  is less than 1\% (Sec. \ref{subsec:Sparse_GEMM_implementation}).


When applying layer-wise sparsity (all supported levels), we observe higher speedup in all cases, \emph{e.g.,} 2.35$\times$ in DeiT-B.
In this case, for DeiT-B, accuracy degradation is still negligible, \emph{e.g.,} 0.12\%. 
However, we observe higher degradation for DeiT-S (1.61\% for int8) and ConvNeXt-S (0.69\% for int8), since they are both small models. 
When applying higher sparsity (uniform 1:3 and 1:4), accuracy degrades even further, but higher speedup is attained, showcasing the accuracy-speedup trade-off.
If we constraint the accuracy degradation to be $\sim$1\% (acceptable in the vast majority of applications \cite{Accuracy_degradation_1_percent_NEURIPS2020}), the optimal speedup is attained for different sparsity configurations in DNNs (bolded in Table \ref{tb:DNN_sparsity_results}).
Specifically, for DeiT-S, uniform 2:4 is optimal (\textbf{1.88$\times$} speedup), while for DeiT-B, uniform 1:4 shows best results (\textbf{3.52$\times$} speedup).
For ConvNeXt-S, layer-wise sparsity (all supported levels) exhibits optimal results (\textbf{2.07$\times$} speedup).
However, when allowing higher accuracy degradation (for instance, within 2\%) higher speedup can be attained, \emph{e.g.,} 3.63$\times$ for uniform 1:4 in ConvNeXt-S.
Finally, notice the higher weight memory reduction as sparsity increases (Table \ref{tb:DNN_sparsity_results}).
Our results demonstrate the importance of supporting multiple sparsity levels in the SST slices.


\vspace{-0.20cm}

\subsection{Versal AIE-ML Sparsity Comparison}
\label{subsec:AIE_ML_comparison}

The new Versal AIE-ML \cite{AMD_AIE_ML_architecture_manual, AMD_AIE_ML_kernel_guide} includes out-of-fabric 
processors that support 2:4 sparsity.
Therefore, we aim to compare the attainable acceleration 
due to sparsity and the efficiency of the compressed sparse format between our SSTs and the AIE-ML.
We exploit AMD's \textit{optimized} codes 
for dense/sparse GEMM kernels for the AIE-ML \cite{AMD_dense_GEMM_int8_github, AMD_AIE_user_guide_2023.2}, using 16-bits for the outputs as this leads to substantially higher compute utilization \cite{AMD_dense_GEMM_int8_github}.
We compile and simulate the AIE-ML designs using Vitis 2023.2 on the VEK280 platform \cite{AMD_VEK_280}.

\begin{figure}[t]
\vspace{-0.40cm}
\centering
\subfloat[]{\includegraphics[width=0.48\linewidth]{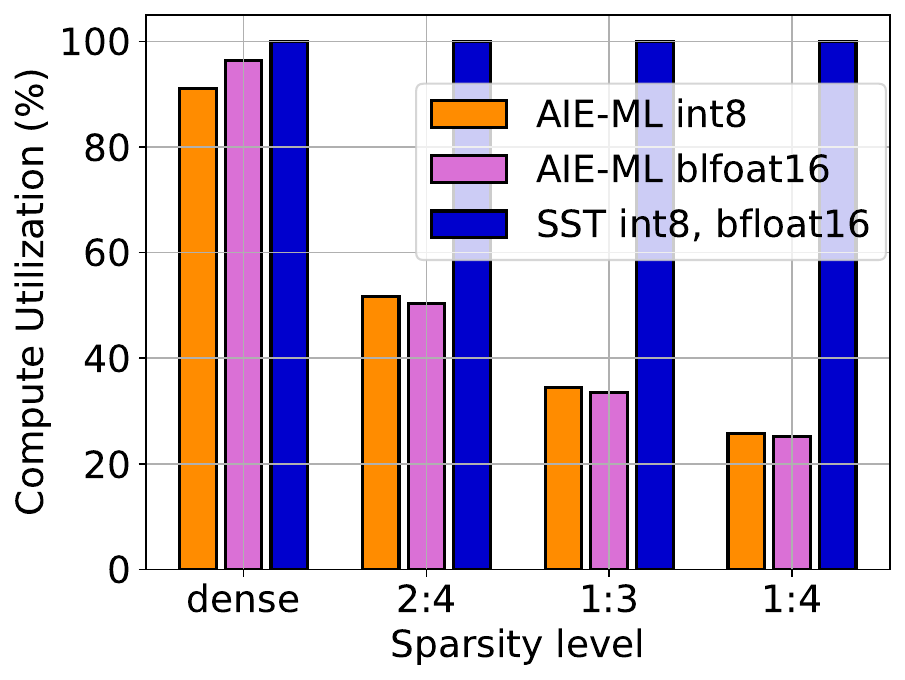}
\label{fig:compute_utilization_SST_vs_AIE_ML}}
\hfill
\subfloat[]{\includegraphics[width=0.455\linewidth]{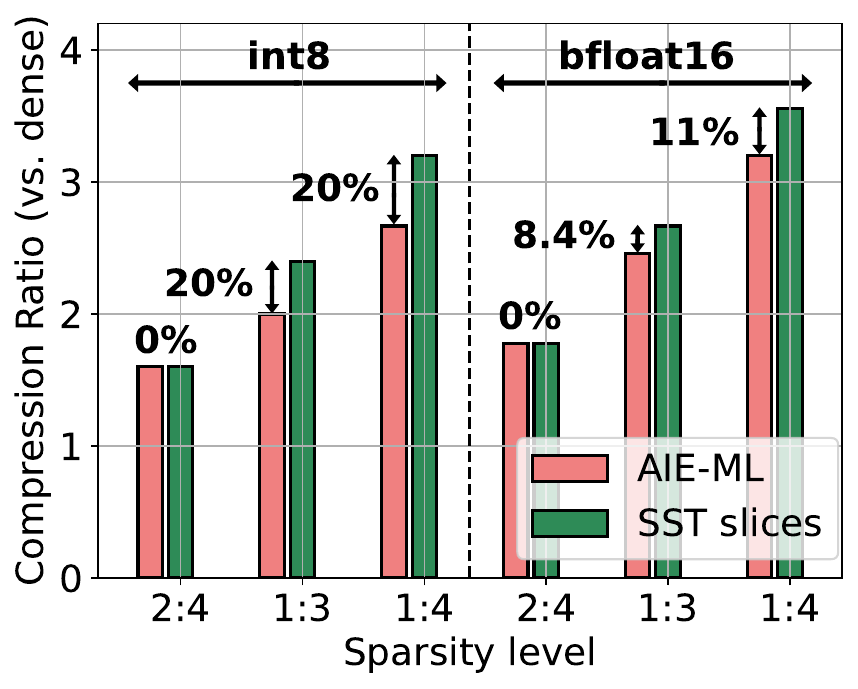}
\label{fig:compression_ratio_SST_vs_AIE_ML}}

\vspace{-0.3cm}

\caption{Compute utilization (a) and compression ratio (b) comparison of a 64$\times$64$\times$64 GEMM using SSTs \emph{vs.} AIE-ML.}
\label{fig:utilization_compres_ratio_SST_vs_AIE_ML}
\vspace{-0.55cm}
\end{figure}

Fig. \ref{fig:compute_utilization_SST_vs_AIE_ML} shows the compute utilization of a 64$\times$64$\times$64 GEMM, mapped to both SST slices and the AIE-ML.
We notice that the AIE-ML achieves higher than 90\% utilization in dense GEMM for both precisions. 
However, for 2:4 sparsity, we observe a substantially lower utilization, \emph{i.e.,} 51.6\% for int8 and 50.4\% for bfloat16.
We note that the compute utilization is calculated as the \textit{effective} utilization of only non-zero computation in the sparse case.
Thus, we obtain a marginal speedup of 13\% (int8) and 4.5\% (bfloat16), for the 2:4 sparse GEMM over the dense case for the AIE-ML.
Moreover, for higher sparsity, \emph{i.e.,} 1:3 and 1:4, the compute utilization in AIE-ML drops more significantly (Fig. \ref{fig:compute_utilization_SST_vs_AIE_ML}).
While higher than 2:4 sparsity can be stored in a compressed format in the AIE-ML, only 2:4 sparsity is inherently supported.
Consequently, no further acceleration can be achieved, leading to significantly decreased \textit{effective} utilization.

This marginal speedup due to sparsity in the AIE-ML is primarily attributed to limited vector load/store bandwidth, causing the sparse GEMM to become I/O bound
\cite{AMD_dense_GEMM_int8_github, AMD_AIE_ML_architecture_manual}.
In contrast, as depicted in Fig. \ref{fig:compute_utilization_SST_vs_AIE_ML}, our SST-based GEMM achieves 100\% utilization for all supported sparsity levels.
This is because we architect our SST slices in a fashion that guarantees the corresponding speedup of each sparsity level, \emph{e.g.,} 4$\times$ for 1:4 sparsity (refer to Sec. \ref{subsec:Sparse_Processing_Element} for design details).

Fig. \ref{fig:compression_ratio_SST_vs_AIE_ML} shows the compression ratios achieved by the SSTs \emph{vs.} the AIE-ML.
The AIE-ML uses a bitmap-based compressed format
\cite{AMD_AIE_ML_kernel_guide}, which is also used in various works accelerating structured sparsity \cite{S2TA_HPCA_2022, STA_arxiv_2020, N_M_sparse_transformers_FPGA_VLSI_2022, Fine_grained_Neural_ODE_FPGA_2023, LAMPS_FCCM_2024}.
Instead, we utilize an index-based format (Sec. \ref{subsec:Fine_grained_structured_sparsity}).
For 2:4 sparsity, both approaches present the same compression ratio (Fig. \ref{fig:compression_ratio_SST_vs_AIE_ML}).
However, for 1:3 and 1:4, our approach leads to higher compression ratio, up to \textbf{20\%} for int8 and \textbf{11\%} for bfloat16, showcasing the storage efficiency of the index-based format.
Moreover, as shown in \cite{N_M_sparse_transformers_FPGA_VLSI_2022}, bitmap achieves higher compression ratio over traditional sparse formats, \emph{e.g.,} CSR, CSC, COO \cite{Sparsity_Hoefler_2021}, for sparsity levels of 50--87.5\%.
This highlights that the supported index-based sparse format provides superior compression over other formats.

\subsection{Impact on Non-AI Benchmarks}
\label{subsec:Non_AI_benchmarks}

We demonstrate the flexibility of our enhanced FPGA architecture with SST slices, by 
exploring the maximum attainable frequency of non-AI benchmarks.
We select non-AI benchmarks from the VTR benchmark suite \cite{VTR_8_2020}, which target various application domains, \emph{i.e.,} $sha$, $mmcl$, $arm\_core$, $stereovision2$, $LU32PEEng$, $bgm$, $raygentop$, and $blob\_merge$  (see \cite{VTR_benchmarks} for domains).
When comparing our enhanced FPGA with a traditional FPGA on the aforementioned benchmarks, we observe, on average, a negligible decrease (\textbf{<1\%}) in maximum frequency. 
Hence, our proposed FPGAs maintain the performance and flexibility of traditional FPGAs, while offering highly efficient sparse and dense DNN acceleration (Sec. \ref{subsec:Dedicated_wires_benefits}--\ref{subsec:Performance_estimation_DNNs}).

\subsection{Insights \& Discussion  }
\label{subsec:Discussion_insights}

\subsubsection{Very High Sparsity in DNN Layers} In the previous sections, we exhibit the importance and efficiency of our proposed in-fabric SST slices for accelerating sparse DNNs.
However, our SSTs support up to 75\% (1:4) structured sparsity.
Prior work \cite{EIE_ISCA_2016, Song_Han_learning_2015} has shown that a few DNN layers can have very high \textit{random} sparsity, \emph{e.g.,} >90\%. 
In this case, \textit{unstructured} sparse FPGA accelerators mapped to traditional CLB and DSP resources, \emph{e.g.,}
\cite{Fowers_sparse_FCCM_2014, Abhishek_sparsity_FPL_2020, unstructured_sparsity_CNN_FPGA_FCCM_2019}, can be exploited to efficiently compute these few high-sparsity layers.
All other DNN layers with less sparsity can be mapped to our SSTs (Sec. \ref{subsec:Sparse_GEMM_implementation} \& \ref{subsec:Performance_estimation_DNNs}), enabling both \textit{structured} and \textit{unstructured} sparse FPGA accelerators to operate synergistically, for maximized efficiency.
Moreover, for very high sparsity (>90\%), the CSR, CSC, COO formats utilized in unstructured sparse accelerators, offer the most efficient compression \cite{Sparsity_Hoefler_2021}.
Instead, for lower sparsity, the index-based format utilized in SSTs is the most efficient (Sec. \ref{subsec:AIE_ML_comparison}).
This emphasizes the custom flexibility of the enhanced FPGAs with in-fabric SST slices in supporting highly efficient sparse DNN accelerators.

\subsubsection{Efficient In-Fabric Slices} 
In-fabric slices have attained commercial success in 
AI-optimized FPGAs, \emph{e.g.,} the Intel tensor blocks \cite{Stratix_10_NX_FPGA_2021} and the new Intel AI-enhanced DSPs \cite{Sergey_Intel_TB_Agilex_5_FCCM_2024, Intel_Agilex_5_tensor_blocks}.
Moreover, recent research \cite{Versal_vs_Stratix_FCCM_2024} comparing leading in-fabric (Intel Stratix 10 NX) with out-of-fabric (AMD Versal ACAP) AI-optimized FPGAs, has shown higher energy efficiency in GEMM for in-fabric solutions. 
However, these efficient in-fabric slices primarily target dense AI acceleration, rendering them insufficient for most DNNs, which are sparse.
Instead, our proposed flexible SST slices support both dense and sparse computation (multiple sparsity levels), showcasing substantial advantages 
over dense in-fabric blocks and traditional FPGAs (Sec. \ref{subsec:Sparse_GEMM_implementation}), for actual sparse DNN workloads (Sec. \ref{subsec:Performance_estimation_DNNs}).
Note that although we are inspired by prior academic research that utilize dense 2D systolic in-fabric blocks \cite{TS_Aman_FPGA_2021, Aman_TS_TRETS_2022}, the methodology described in Sec. \ref{sec:Architecture_Overview} for introducing sparsity support can be generalized in straightforward fashion to other in-fabric FPGA blocks.

\subsubsection{Future Directions} 
Our SST slices support \textit{static} sparsity targeting weights in DNNs.
Future work could also explore supporting \textit{dynamic} sparsity in activations.
The SST slices could also enable other modes for increased flexibility, \emph{e.g.,} element-wise modes, similar to \cite{TS_Aman_FPGA_2021, Aman_TS_TRETS_2022}.
However, this is beyond the scope of this paper and is therefore left for future work.
We note that general matrix-vector (GEMV) can be directly supported in SSTs at batch size of four with 100\% utilization (by setting $X$=1 in designs of Sec. \ref{subsec:Matrix_multiplication_mapping}). 
Additionally, multiplexing logic can be incorporated into SSTs to increase GEMV utilization for batch size of one, as shown in \cite{Aman_TS_TRETS_2022}, which we leave as future work.
Finally, another extension is to include multiple input crossbars, which would further reduce the small area overhead of the SSTs, and quantify the trade-offs in area and FPGA routability.



\section{Conclusion}
\label{sec:Conclusion}

Structured sparsity support has been incorporated in state-of-the-art GPUs, \emph{e.g.,} Nvidia H100 and AMD MI300, as well as the new Versal AIE-ML processors.
Although FPGA architectures have been enriched with in-fabric blocks for DNN acceleration, these blocks are primarily designed for dense operation.
This leads to insufficient computation for most contemporary DNN models, which display varying levels of sparsity.
To effectively address this challenge, we propose flexible in-fabric blocks, named SST slices, that support multiple levels of structured sparsity. 
We show that our sparse GEMM accelerators exploiting the SST slices offer substantial performance, scalability, and area advantages over traditional FPGAs. 
Demonstration on various state-of-the-art sparse DNN models utilizing our SST slices, exhibits up to 3.52$\times$ speedup with marginal accuracy degradation compared to dense in-fabric acceleration.

\vspace{-1.2mm}

\section*{Acknowledgment}
This work was supported by the National Science
Foundation CCF Grant No. 2107085, the ONR Minerva program, and iMAGiNE -- the Intelligent Machine Engineering Consortium at UT Austin.


\newpage

\bibliographystyle{ACM-Reference-Format}
\bibliography{bibliography}

\end{document}